\documentclass[journal=jacsat,manuscript=article]{achemso}

\usepackage[version=3]{mhchem} 
\usepackage{footnote}
\usepackage{graphicx}
\usepackage{epstopdf}
\usepackage{bm}
\usepackage{mathtools}
\usepackage{textcomp}
\usepackage{amsmath}
\usepackage{physics}
\usepackage{float}
\usepackage{siunitx}
\usepackage{amssymb}
\usepackage{float} 
\usepackage{steinmetz}
\usepackage[utf8x]{inputenc}
\usepackage{csquotes}
\usepackage{lipsum} 
\usepackage{soul,color}
\usepackage{seqsplit}

\author{Anum Nisar}
\affiliation[Monash University]
{ARC Centre of Excellence in Exciton Science and School of Chemistry, Monash University, Clayton, Victoria, 3800, Australia}
\author{Harini Hapuarachchi}
\affiliation[RMIT]
{ARC Centre of Excellence in Exciton Science and Chemical and Quantum Physics, School of Science, RMIT University, Melbourne, 3001, Australia}
\author{Laurent Lermusiaux}
\affiliation[Monash University]
{ARC Centre of Excellence in Exciton Science and School of Chemistry, Monash University, Clayton, Victoria, 3800, Australia}
\altaffiliation
{Current address: Université de Lyon, CNRS, École Normale Supérieure de Lyon, Laboratoire de Chimie UMR 5182, 46 allée d’Italie, F-69007 Lyon, France.}
\author{Jared H. Cole}
\affiliation[RMIT]
{ARC Centre of Excellence in Exciton Science and Chemical and Quantum Physics, School of Science, RMIT University, Melbourne, 3001, Australia}
\email{jared.cole@rmit.edu.au}
\author{Alison M. Funston}
\email{alison.funston@monash.edu}
\affiliation[Monash University]
{ARC Centre of Excellence in Exciton Science and School of Chemistry, Monash University, Clayton, Victoria, 3800, Australia}

\title[Hybrid Assemblies]
  {Enhanced Control of Quantum Dot Photoluminescence in Hybrid Assemblies}

\keywords{Add keywords here}

\begin{document}

\begin{tocentry}

Some journals require a graphical entry for the Table of Contents.
This should be laid out ``print ready'' so that the sizing of the
text is correct.

Inside the \texttt{tocentry} environment, the font used is Helvetica
8\,pt, as required by \emph{Journal of the American Chemical
Society}.

The surrounding frame is 9\,cm by 3.5\,cm, which is the maximum
permitted for  \emph{Journal of the American Chemical Society}
graphical table of content entries. The box will not resize if the
content is too big: instead it will overflow the edge of the box.

This box and the associated title will always be printed on a
separate page at the end of the document.

\end{tocentry}

\begin{abstract}
 The distance-dependent interaction of an emitter with a plasmonic nanoparticle or surface forms the basis of the field of plexitonics. Semiconductor quantum dots (QDs) are robust emitters due to their photostability, and, as such, offer the possibility of understanding the fundamental photophysics between one emitter and one metal nanoparticle. Hence, a key enabling challenge is the formation of such structures (\textit{i.e.} systems containing both QDs and plasmonic nanoparticles) in high purity. Here, we present the translation of DNA-based self-assembly techniques to assemble metal and semiconductor nanocrystals into discrete hybrid structures, including dimers, of high purity. This method gives control over the interparticle separation, geometry, and ratio of QD:metal nanoparticle, as well as the spectral properties of the metal/QD components in the assembly to allow detailed investigation of plasmon-exciton interaction. The hybrid assemblies show the expected enhancement in steady-state photoluminescence accompanied by an increase in the QD emission rate for assemblies with a strong overlap between the QD emission and localised surface plasmon resonance. In contrast, lengthening of the QD emission lifetime (a reduction of the emission rate) of up to 1.7-fold, along with an enhancement in steady-state PL of $\sim$15 - 75 \% is observed upon detuning of the QD and metal nanoparticle spectral properties. These results are understood in terms of the Purcell effect, where the gold nanoparticle acts as a damped, nanoscale cavity. Considering the metal nanoparticle using generalised nonlocal optical response theory (GNOR) and the QD as an open quantum system, we show that the response is driven by the interference experienced by the emitter for parallel and perpendicular field orientations. This understanding provides a mechanism for control of the emission rate of a QD by a metal nanoparticle across a much wider range of lifetimes than previously understood.
\end{abstract}

\section{Introduction}

The excitonic states and thus the optical properties of light emitters can be controlled by judicious engineering of their local environment.\cite{Bellessa_2004_PhysRevLett} One method to modify the electromagnetic properties of emitters is via the Purcell effect, whereby the photonic environment (\textit{i.e.}the dielectric environment) of the emitter is controlled or changed. In this vein, optical microcavities, photonic crystals, interfaces, and gratings have all been used to control the emission rate of the fluorophores.\cite{Bayer_2001_PhysRevLett,santhosh_2016_naturecomm} However, these approaches require precise control of the position and orientation of optically active materials at the nanoscale.

Metal nanoparticles that are much smaller than the wavelength of incident light ($\lambda$) exhibit strong resonant excitations known as localized surface plasmon resonances (LSPRs) \cite{hapuarachchi2018optoelectronic, liu2017poly}. LSPRs are non-propagating modes of excitation of the conduction band electrons arising naturally from the scattering problem of a subwavelength metal nanoparticle in an oscillating electromagnetic field \cite{maier2007plasmonics}. These excitations enable metal nanoparticles to act as nanoscale optical cavities, able to focus electromagnetic energy to spots much smaller than $\lambda$ overcoming the half-wavelength size limitation of conventional optical cavities \cite{hapuarachchi2017cavity, ridolfo2010quantum, waks2010cavity, gettapola2019control}. Thus, metal nanoparticles can significantly influence the optical properties of quantum emitters, such as nanocrystal quantum dots, placed in nanoscale proximity \cite{hapuarachchi2018exciton}.

Interaction between an emitter and a plasmonic metal nanoparticle gives rise to phenomena including plasmon-enhanced fluorescence,\cite{Lee_2004_Nanolett,Munechika_2010_Nanolett} surface-enhanced Raman scattering,\cite{Hugall_2009_ApplPhysLett} energy transfer,\cite{Hao_2011_AngewChemieIntEd,Quach_2011_JAmChemSoc} and coupling between plasmon and molecular resonances.\cite{Urena_2012_AdvMater} The range of potential applications within which the phenomenon of plasmon-exciton resonance coupling may be exploited is vast and diverse, encompassing plasmon-enhanced catalytic reactions,\cite{zheng_2007_InorgChem,Mahanti_2014_ChemPhysLett} plasmon-enhanced optical signals,\cite{Mahanti_2014_JLumin,Peh_2010_MaterLett} solar cells,\cite{Gan_2013_AdvMater,Schladt_2010_AngewChemieIntEd} and biotechnological applications.\cite{Zhou_2012_JMaterChem} Various factors affect the plasmon-exciton coupling. Metal nanoparticles modify the electric field at their surface, thereby influencing the dielectric environment of nearby emitters and consequently their emission (intensity and emission rate). Another mode of interaction is damping which occurs due to energy transfer from the emitter to the metal particle acceptor. The competition between field enhancement and nonradiative damping due to energy transfer to the LSPR controls the optical response.\cite{Sajanlal_2011_Nanoreviews} Effective plasmon-fluorophore coupling requires the close proximity of the plasmon and fluorophore and significant overlap of the emission spectrum of the fluorophore with the LSPR of the metal nanocrystals.\cite{Sajanlal_2011_Nanoreviews} Dynamic control of this exciton-plasmon system opens up new avenues for novel and efficient exciton-based optoelectronic applications.

In these coupled exciton-plasmon systems, semiconductor quantum dots (QDs) offer significant advantages compared to organic fluorophores, such as photostability, high quantum yield, high absorption cross-section, and narrow, tunable emission.\cite{CostaFernandez_2006_TrAC,Baskoutas_2006_Jappliedphysics} Consequently, hybrid structures incorporating quantum emitters and metal nanoparticles are promising candidates for use across the range of potential applications exploiting their interactions.

Hybrid plasmon-exciton structures have been realised by the fabrication of QD-Au films, consisting of a superposition of Au or QD thin films separated by spacer layers. Within these, both quenching and enhancement of the PL intensity were achieved upon tuning the spectral overlap and variation of the polymer/polyelectrolyte layers between the films.\cite{kim_2019_RSCadv,leblanc_2013_Nanolett,kulakovich_2002_Nanolett}. A major concern for NP-based films is the aggregation of NPs upon drying. This causes changes in the LSPR of the metal NP and the emission of the fluorophore, thereby modifying the overall optical response of the hybrid films compared to their counterparts in the solution.\cite{chaikin_2013_Anal_Chem} Electron beam lithography approaches have also been widely employed to prepare two-dimensional hybrid structures, although gap variation and surface roughness lead to variations in the optical response of the coupled nanoparticles.\cite{Jensen_2000_JPhysChemB,Ratchford_2011_Nanoletters,ratchford_2011_NanoLett} 

The strong exciton-plasmon coupling has been observed in controlled, discrete systems with a one-to-one stoichiometry for the QD and metal nanoparticle. Precise profiling of distance-dependent fluorescence quenching and enhancement between a single gold nanoparticle and a quantum dot has been carried out via AFM-based micromanipulation\cite{eckel_2007_small, masuo_2016_ACSPhotonics} and by the incorporation of a single QD in a plasmonic cavity\cite{chikkaraddy_2016_Nature,leng_2018_NatureComm} in proof-of-principle experiments. In contrast, colloidal methods have the potential for the development of three-dimensional structures in scalable quantities. Energy transfer in electrostatically assembled clusters containing multiple dyes or quantum dots around metal nanoparticles showed a 2 - 3-fold reduction in the lifetime of the fluorophore, whilst the transfer efficiency was able to be modulated from 30\% to 55\% via a change in the concentration of either gold nanoparticles or the fluorophore\cite{sen_2007_AppPhyLett,wargnier_2004_NanoLett} However, such approaches do not provide control over the number of emitters or interparticle separation. To overcome this, chemical coupling of the metal and fluorophore with short biological or chemical linkers, or silica encapsulation have been employed to develop hybrid structures, \cite{Kang_2005_AngewandteChemieInternalEdition,Ofir_2008_Chemsocreviews,Quintiliani_2014_JMaterialschemC,Lin_2016_ACSnano} although generally, these methods provide only limited control over metal-fluorophore distances and stoichiometry, limiting their use. Consequently, a key requirement in this field is the development of assembly methods with the ability to fully control the relative location, orientation, and interparticle separation in a coupled plasmon-exciton systems.  

DNA has been exploited to achieve self-assembly of gold nanoparticles by hybridisation of single-stranded DNA attached to metal nanoparticles as well as sophisticated DNA origami scaffolds.\cite{li2009nanofabrication,schreiber_2014_Nature,wang_2012_Angewandte,zhang2019dna} The variable length of the DNA strands offers flexibility in the interparticle separation, and purification techniques enable control over the number of particles in the assembly.\cite{liu2018dna,guerrini2012tuning} Alivisatos pioneered the DNA-based self-assembly of gold nanoparticles into discrete structures.\cite{alivisatos_1996_Nature} Subsequently, improved protocols for DNA-based gold nanoparticle assembly have been developed.\cite{fu_2004_JACS} In contrast, DNA-based QD assembly protocols are relatively underdeveloped,\cite{tikhomirov_2011_Nature,Lermusiaux_2020_NanoRes} with a key enabling capability, the purification of particles with one (or a given number) of attached DNA strands per particle lacking. Assembly of DNA functionalized CdTe QDs into nanoassemblies has been shown, although the purification technique (size-selective filtration) does not allow purification of QDs with only one or two DNA strands per particle.\cite{tikhomirov_2011_Nature} Gel electrophoresis approaches have also been used to purify DNA based hybrid Au/QD assemblies\cite{fu_2004_JACS,Lermusiaux_2020_NanoRes,fernandez2020hybrid}. To date, visualisation of QDs in DNA-based QD assemblies (lacking defined numbers of DNA strands/particle) has been achieved using fluorescent straining which interferes with the optical signal of the QDs. The formation of nanostructures that show strong exciton-plasmon coupling requires two different materials, with different surface chemistries and sizes, to be assembled together in high yield, with control over geometry. The additional challenges associated with the incorporation of different materials in the assembly are significant. 

The optical properties of metal-fluorophore systems are somewhat incongruous and vary between quenching\cite{vaishnav2018long,pons2007quenching} and enhancement\cite{tripathi2014long,tripathi2013plasmonic} of the photoluminescence intensity of the emitter. Within discrete hybrid assemblies with a good overlap of the metal LSPR and fluorophore emission, strong quenching of the QD PL intensity is observed,\cite{pons2007quenching,nikoobakht2002quenching} but at a given separation distance and with a decrease in the spectral overlap, a cross-over from quenching to enhancement in the PL intensity occurs.\cite{haridas2011photoluminescence,huang_2015_Nanoscale,Liu_2006_JACS,inoue_2015_ACS} Photoluminescence enhancement has, however, also been observed in highly resonant Au-QD systems. For example, 2.5 to 8-fold photoluminescence enhancements, with little to no change in photoluminescence lifetimes, for on-resonant Au-QD assemblies with interparticle separations in the range of $2-\SI{20}{\nano\meter}$ have been reported.\cite{kormilina2018highly,ribeiro2013enhanced,nepal_2013_ACS} Higher degrees of steady-state photoluminescent enhancement have been observed for QD-Au hybrid films where increases of up to 50-fold have been demonstrated.\cite{hsieh2007mechanism,song2005large} In contrast to the ambiguity in the PL intensity of hybrid systems, PL decay measurements show a reduction in the lifetime of the emitter close to the metal for all the reported hybrid systems, irrespective of quenching and enhancement in the PL intensity \cite{li2016spectral,Ratchford_2011_Nanoletters}. This will be further discussed in our theoretical insights section.

Here, we report an approach that overcomes many synthetic challenges to allow the controlled organization of quantum emitters near gold nanoparticles with control over emitter wavelength, metal LSPR, interparticle spacing ($\SI{10}{\nano\meter}$ and $\SI{34}{\nano\meter}$), the number of emitters and their location around the metal nanoparticle. The assemblies were achieved with high colloidal stability, reproducibility, and purity. We demonstrate the modulation of the CdSe QD lifetime by the AuNP via the tuning of different parameters. The hybrid assemblies show increased control over the QD lifetime, with an increase in the lifetime observed when the QDs and LSPR of the metal are highly detuned, along with the expected shortening of the lifetime when these are resonant. The photoluminescence across this range consistently shows an increase in intensity. We theoretically demonstrate that the observed decay rate modifications are likely to stem from the cavity-like regulation of the QD local electromagnetic environment by gold nanoparticles. Furthermore, by modelling single metal-QD dimers using an open quantum systems framework, we demonstrate the high orientation dependence of their photoluminescence. The origin of PL intensity enhancements observed experimentally in our colloidal ensembles can be attributed to the dominance of axial field enhancements.

\section{Results and Discussion}

In this work, we synthesise discrete assemblies of AuNPs and CdSe/CdS QDs using a DNA-driven self-assembly approach.  Assemblies are formed by mixing two sets of particles functionalised with complementary single-stranded DNA as shown in Scheme \ref{Hybridization_Scheme1}. The desired assemblies are made of gold nanospheres (10 nm diameter - AuNS$_{10}$) or nanorods surrounded by one (hybrid dimers) or several QDs (core-satellite). To obtain the hybrid nanostructures in high yield, an electrophoretic method to obtain QDs attached to exactly one DNA strand per particle, while retaining the fluorescence quantum yield is developed. This ability expands the currently available library of assembled nanostructures able to be achieved, with the possibility for the selective incorporation of semiconductor materials into already established DNA-based structures.

The number of QDs attached to the gold core is controlled by varying the number of DNA strands adsorbed onto the AuNPs. The distance between the AuNPs and QDs is given by the DNA length. Two different sets of DNA strands which are 34 and 100 bases long (theoretical distances of 10 and 34 nm) are used, as is a final purification step, again with electrophoresis, to obtain the desired products in high purity. Overall, this DNA-based electrophoretic approach allows control over the number of emitters in the hybrid assembly, the interparticle separation between AuNP and QDs and the nanoparticle size and shape.

\subsection{Building Blocks for Self-Assembly}

The transferable nature of the DNA functionalization across different sized nanoparticles of the same material (and therefore with the same surface chemistry\cite{Lermusiaux_2020_NanoRes}) allows the incorporation of quantum dots of different sizes, and therefore different energy band gaps, into the assemblies using the same protocols. The QDs used have diameters of 3.5$\pm 1.3$ nm (Q550), 4.5$\pm 0.8$ nm (Q570), 6.0$\pm 0.6$ nm (Q610), and 8.0 $\pm 1.0$ nm(Q650), with the wavelength of their lowest energy absorption 545 nm (Q550), 548 nm (Q570), 589 nm (Q610), and 622 nm (Q650).  

The metal nanoparticles and quantum dots are synthesized according to standard protocols\cite{piella_2016_ChemMat,nikoobakht_2003_ChemMat,carbone_2007_Nanolett,boldt_2013_ChemMat} and electron microscopy images of these are shown in Figure S1. The synthesis of quantum dots was carried out in organic solvent to achieve narrow particle size distribution and high photoluminescence. The former is required to achieve well-separated, discrete bands on the agarose gel employed in the DNA-based assembly approach, while the latter is required for visualisation of the separated particles. Phase transfer of the QDs into an aqueous solution is achieved with PsAA allowing good retention of the QD PL. Metal nanoparticles are ligand exchanged with \textit{bis}(\textit{p}-sulfonatophenyl)phenylphosphine (BSPP). We refer to the gold nanospheres as AuNS$_{10}$ and gold nanorods as AuNR$_{2.7}$.

\subsection{Functionalisation of QDs and Au Nanoparticles with DNA}

To enable the formation of the desired discrete assemblies in high purity, QDs and gold nanoparticles attached to single complementary DNA strands are prepared and mixed as shown in Scheme \ref{Hybridization_Scheme1}.

\begin{scheme}[!ht]
  \includegraphics[width=\textwidth]{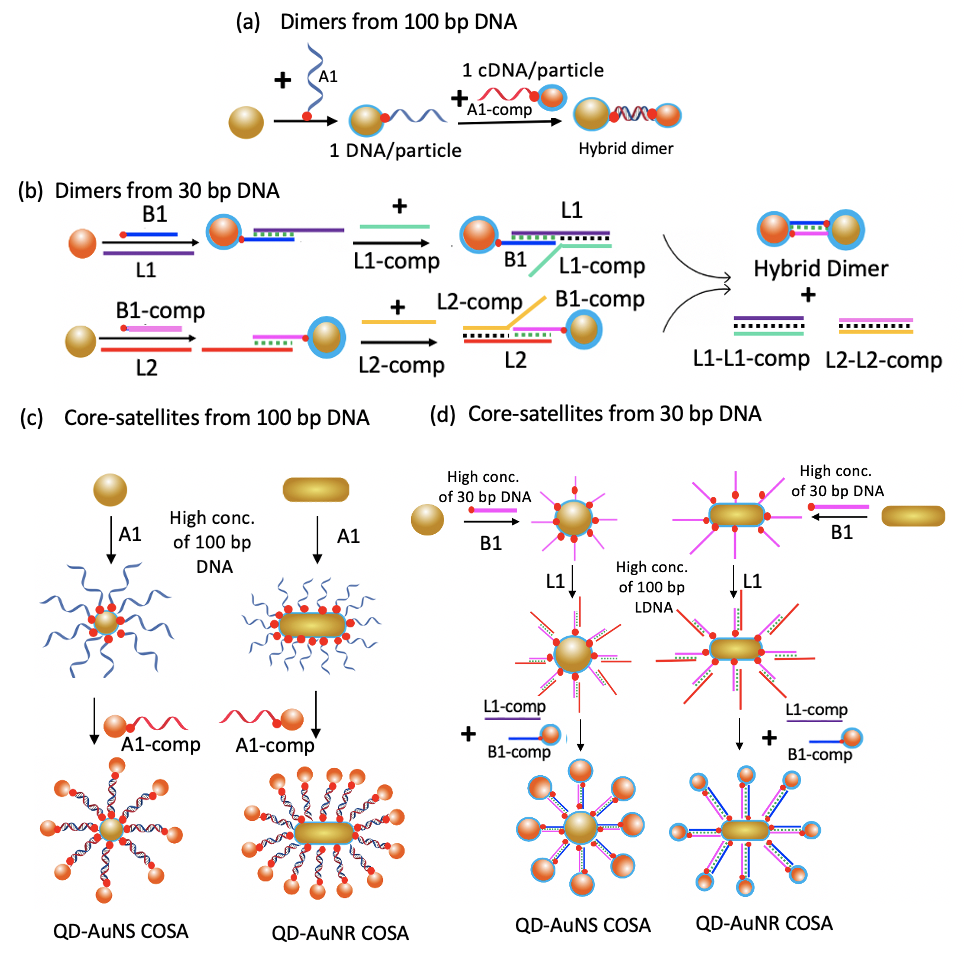}
  \caption{Formation of hybrid structures with different strand lengths of DNA. The wiggly lines and straight lines represent 100 and 30 bp DNA, respectively. The orange sphere represents a quantum dot while the golden sphere and rod are representing gold nanosphere and nanorod. (a) Hybrid dimers with long DNA (100 bp), $\sim\SI{34}{\nano\meter}$ interparticle separation. (b) Hybrid COSA with long DNA (100 bp), $\sim\SI{34}{\nano\meter}$ interparticle separation. (c) Hybrid dimers with short DNA (30 bp), $\sim\SI{10}{\nano\meter}$ interparticle separation. (d) Hybrid COSA with short DNA (30 bp), $\sim\SI{10}{\nano\meter}$ interparticle separation.}
  \label{Hybridization_Scheme1}
\end{scheme}

To perform DNA functionalisation of the QDs with 100-mer ssDNA, a concentrated solution of ligand-exchanged QDs ($\sim$25 $\mu$M) were incubated with 100 bp A1-comp tri-thiolated DNA strands, at a concentration optimized for attachment of one strand of DNA per particle ($\sim$10-40 $\mu$M for QDs) in salt for 24 hours to allow adsorption of the thiol groups to the nanoparticle surface. The concentration of the QDs used here is higher than that of the gold nanoparticles to allow their visualisation in the gel, however, the relative ratio of DNA to nanoparticles required for the QDs is lower compared to that for the gold nanoparticles. The presence of salt reduces the electrostatic repulsion between the particle and DNA strands, allowing the DNA to bind to the nanoparticle surface.\cite{busson_2011_NanoLett} Following this, to improve the colloidal stability and photoluminescence quantum yield, QDs were passivated with mercapto(ethylene glycol)$_{6}$ carboxylate. PEGylation also promotes the dispersion of QDs in the high salt-containing buffer media required for both electrophoresis and DNA hybridisation.

\begin{figure}[!ht]
    \includegraphics[width=0.95\textwidth]{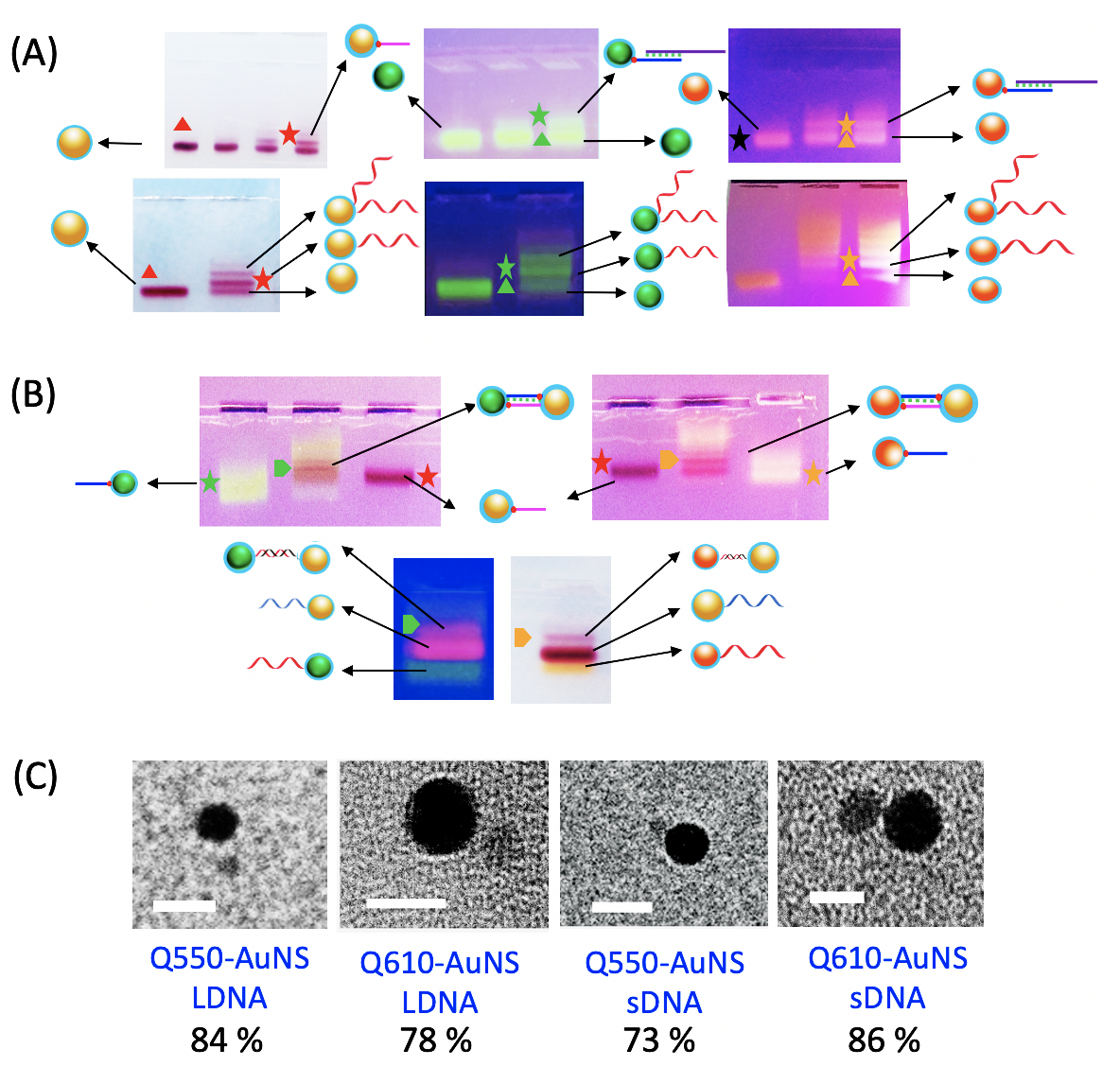}
\caption{(A) Gel images showing the purification of AuNS$_{10}$, Q550 and Q610 with defined numbers of DNA strands per particle using agarose gel of 2.75\%, 4.1\%, and 3.4\%, respectively, for 35-40 min at $\SI{80}{\volt}$. The AuNS$_{10}$, Q550, and Q610 are represented as golden, green, and orange coloured spheres respectively. The curved and straight lines indicate the attachment of long DNA (LDNA) and short DNA (sDNA), respectively. The DNA concentration increases from left to right in the gel figures. The fastest band contains nanocrystals with no DNA; AuNS (red triangle), Q550 (green triangle), and Q610 (orange triangle). The second band is nanocrystals with 1 DNA/particle i.e. AuNS$_{10}$-1DNA (red star), Q550-1DNA (green star), and Q610-1DNA (orange star). (B) Purification of the hybrid Q550-AuNS$_{10}$ (green diamond) and Q610-AuNS$_{10}$ dimer assemblies (orange diamond) from the unconjugated AuNS$_{10}$-1DNA (red star) and Q550-1DNA (green star) and Q610-1DNA (orange star) particles after running on 2.5\% agarose gel at $\SI{80}{\volt}$ for 30 min. (C) Representative TEM images of the hybrid Q550-AuNS$_{10}$ and Q610-AuNS$_{10}$ dimers synthesized with 30 and 100 bp of DNA. Scale bar is 10 nm}
\label{Dimers_Gels}
\end{figure}

Electrophoretic separation of QDs attached to different numbers of DNA strands is carried out by optimisation of the salt concentration and gel density, and using different DNA concentrations (Figure \ref{Dimers_Gels}A). Well-separated, clear bands were observed upon electrophoretic separation of the DNA functionalised QDs, indicating separation of samples containing discrete numbers of DNA strands per particle. The number of DNA strands attached to the nanoparticles is determined by a comparison of each band position with respect to the reference band of the nanoparticles without DNA. An increase in DNA concentration leads to the monomer band (without DNA) disappearing, along with the appearance of higher bands indicative of QDs with multiple DNA strands per particle, including one, two, and three strands/particle. The agarose gel concentration ranges from 3.2\% - 4.1\% depending on the size of the QDs. This is relatively high compared to that used for gold nanoparticles (2.75\%) due to the small size of the QDs compared to the gold nanoparticles. The DNA-functionalised QDs were bright enough to be visualised in the gel from their luminescence (Figure \ref{Dimers_Gels}A) due to the relatively good retention of their quantum yield, without the use of staining agents such as ethidium bromide. Extraction of particles functionalized with one DNA strand per particle was achieved by slicing out the separated band and soaking the gel slice in TBE buffer overnight.

To form assemblies with short DNA strands to realise stronger interaction between the Au and QD, the challenge is to achieve well-resolved gel bands containing one DNA strand per particle on purification via electrophoresis. The short-chain length of the 30 bp DNA does not allow a significant size increase of the DNA-nanoparticle conjugate (compared to the nanoparticle itself), which makes it difficult to purify via electrophoresis. A toehold-mediated DNA displacement approach was thus used to achieve this goal\cite{busson_2011_NanoLett} as outlined in Scheme \ref{Hybridization_Scheme1}B. Different volumes of 30 bp B1 DNA (100 $\mu$M) are added to a concentrated solution of the QDs, along with an equimolar concentration of the 100 bp long non-functionalized lengthening DNA strand (L1). The L1 DNA has 15 bp complementary to the B1 DNA. The B1 DNA attaches the AuNP via trithiol moiety while its overhanging strand hybridises with the L1 DNA. This conjugation increases the effective length/bulk of the DNA, therefore facilitating the separation of the nanoparticles in the gel (Figure \ref{Dimers_Gels}A).\cite{busson_2011_NanoLett} 

The DNA functionalization and separation of products to isolate nanoparticles containing one DNA strand per particle of small gold nanospheres ($D = 10 nm$) is well-known and was carried out according to literature procedures\cite{Lermusiaux_2020_NanoRes,yao_2006_NanoTech,lermusiaux_2018_Nanoscale,chen_2015_ACSPhoton} by incubating gold nanoparticles ($\sim$1-5 $\mu$M) with 100 bp A1 DNA ($\sim$5-30 $\mu$M) and their subsequent purification via gel electrophoresis (shown in Figure \ref{Dimers_Gels}A). The procedure outlined above was repeated for 30 bp DNA with gold nanoparticles to achieve one strand of B1-comp and L2 DNA per gold nanoparticle.\cite{busson_2011_NanoLett} The nanoparticles were then used for the formation of the assemblies.

\subsection{Self-Assembly of Au-QD Dimers}

To form discrete QD:Au hybrid dimers in high yield with the 100 bp DNA linker (Scheme \ref{Hybridization_Scheme1}A), the AuNP and QDs (containing one A1-DNA and one complementary A1-DNA per particle, respectively) are incubated in the presence of salt (Scheme \ref{Hybridization_Scheme1}A). Hybrid Au-QD dimer assemblies with 30 bp linkers are formed by mixing B1 and L1 DNA functionalized QDs with B1-comp and L2 DNA functionalized gold nanoparticles along with L1-comp and L2-comp lengthening DNA strands. These were hybridised overnight, resulting in a branch migration which displaces the stable final duplex (consisting of the non-thiolated 100-mer DNA pairs of L1 - L1-comp, L2 - L2-comp), along with the short DNA bound Au-QD assembly. 

Following incubation of the DNA functionalised nanoparticles, the assemblies were further purified via electrophoresis to remove unhybridised nanoparticles and higher-order assemblies. Figure \ref{Dimers_Gels}B presents the gels (2.5\%) of the electrophoretic purification of the short and long DNA assembled hybrid dimers Q550-AuNS$_{10}$ and Q610-AuNS$_{10}$. The first and second bands (from the bottom) in the gels correspond to the monomer QD/AuNPs functionalized with one DNA strand, respectively, whilst the third band corresponds to the hybrid dimer assembly. No slower bands, indicative of the formation of higher-order assemblies, are observed. The bands containing the assemblies were eluted from the gel and characterized by transmission electron microscopy (TEM) as shown in Figure \ref{Dimers_Gels}C. The high contrast particles in the TEM images are gold nanoparticles, while the nanoparticles with lower contrast are the semiconductor QDs. The dimer yields range from 73\% to 86\% (Figure S3) following deposition on the substrate. The distance between the Au nanoparticle and QD, as calculated from the TEM images, is $1-\SI{1.5}{\nano\meter}$ with 100 bp DNA and $0.5-\SI{1}{\nano\meter}$ with 30 bp DNA. The theoretically calculated interparticle separation between the metal NP and QD with fully-extended 30 bp and 100 bp DNA, as expected in solution phase, is $\sim\SI{10}{\nano\meter}$ and $\sim\SI{34}{\nano\meter}$, respectively. The smaller interparticle separation achieved on deposition compared to the theoretically predicted distance based on the length of the DNA double-stranded helix linker is likely due to the drying effects upon deposition, including van der Waals interactions between the particles and capillary drying effects.  

\subsection{Self-Assembly of Au-QD Core-Satellite (COSA) Structures}

\begin{figure}[ht!]
    \includegraphics[width=0.84\textwidth]{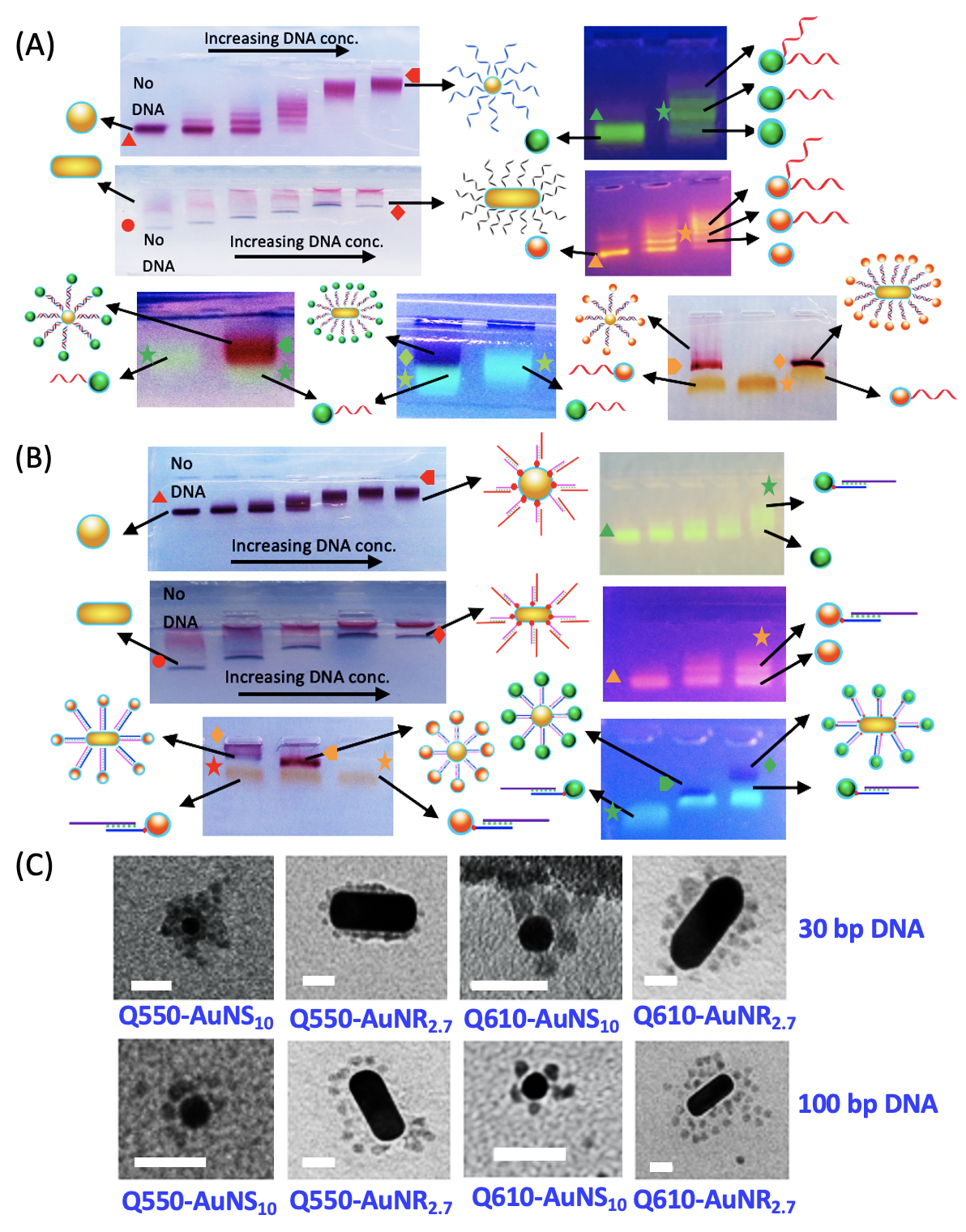}
\caption{Formation and purification of hybrid core-satellite assemblies with (A) Long DNA (LDNA) of 100 bp (curved lines) and (B) Short DNA (sDNA) of 30 bp (straight lines). The DNA concentration increases from left to right with the leftmost band the particle with no added DNA. Gel images show the DNA purification of fully functionalized central AuNS$_{10}$ (red pointer) and AuNR$_{2.7}$ (red diamond) with reference to the particle with no attached DNA; AuNS$_{10}$ (red triangle) and AuNR$_{2.7}$ (red circle) on 1\% agarose gel at $\SI{80}{\volt}$ for 35 minutes. Hybrid Q550-AuNS$_{10}$ COSA (green pointer) and Q610-AuNS$_{10}$ COSA (orange pointer) assemblies were purified from the unconjugated Q550-1-DNA (green star) and Q610-1DNA (orange star) particles by running on 1.5\% agarose gel at $\SI{80}{\volt}$ for 30 minutes. (C) TEM characterization of the synthesised hybrid COSA assemblies of Q550 and Q610 with 30 and 100 bp DNA. The scale bar is 20 nm.}
\label{COSA_gels}
\end{figure}

The versatility of the DNA-based approach following isolation of QDs containing one DNA strand per particle is highlighted by the adaptation of the DNA hybridisation scheme to form Au-QD core-satellite structures. The fabrication of core-satellite hybrid structures (Scheme \ref{Hybridization_Scheme1}C) containing a metal core and multiple monofunctionalised QD satellites were formed by initially conjugating the central gold nanoparticle with an excess of DNA prior to the self-assembly. The higher DNA density is achieved by mixing the particles with higher DNA (40 - 90 $\mu$M) and salt concentrations (50 - 100 $\mu$M), and is confirmed by lower electrophoretic mobility of the particles on the agarose gel (Figure \ref{COSA_gels}A). The high salt concentration reduces the electrostatic interaction between the negatively charged ssDNA and thus allows a higher DNA density to be functionalised onto the  AuNP surface. Additionally, running the cores on the gel separates the DNA-coated particles from excess (non-binding) DNA strands and is a necessary step prior to assembly. It also has the benefit of purifying the desired nanoparticles from side-products (different shapes for example) as can be seen in the gel images for the purification of AuNR, which are separated from nanosphere byproducts, the blue and red bands respectively in the gels in Figure \ref{COSA_gels}A ($\sim\SI{50}{\nano\meter}$ AuNS$_{50}$).  

To form hybrid QD-AuNS$_{10}$ COSA and QD-AuNR$_{2.7}$ COSA assemblies with 100 bp DNA, the fully A1 DNA functionalised AuNS$_{10}$ and AuNR$_{2.7}$ core particles are hybridised with QD satellite particles containing one complementary A1-comp DNA strand per particle. The hybrid core-satellite assemblies were again purified via a second gel electrophoresis step on a diluted gel (1.5\%) to separate the hybrid COSA assemblies from the unbound satellites (Figure \ref{COSA_gels}B and S8). The low mobility of the COSA assembly compared to single unconjugated nanoparticles results in separate bands for these species as shown in Figure \ref{COSA_gels}A (lower panel). Similarly, to form 30 bp DNA-based COSA structures, a high concentration (40-90 $\mu$M) of B1-DNA and L1 lengthening strand are mixed with the core AuNP, which, after purification, is hybridized with QDs functionalised with one complementary B1-comp and L1-comp DNA duplex per particle as outlined in (Scheme \ref{Hybridization_Scheme1}D).

\clearpage

The TEM images of the COSA structures in Figure \ref{COSA_gels}C and S3 show the distribution of the number of satellites per metal core ranges from 1 - 10 for AuNS$_{10}$ and 1 - 20 for AuNR$_{2.7}$ hybrid assemblies (Fig S4, Table \ref{Lifetimes}). As expected, the number of satellites around the metal core decreases as the size of the QD increases from Q550 to Q650. The coverage of the satellites around the core NP also depends upon the available surface area of the core NP. Therefore, a higher DNA valency, and therefore a higher number of QD satellites, is observed for gold nanorod cores AuNR$_{2.7}$ relative to gold nanospheres of 10 nm diameter. 

\subsection{Photochemistry}

The 1:1 metal:QD assemblies allow the effect of the presence of just one metal nanoparticle on a QD to be quantified, and the experimentally observed optical response accurately modelled to understand the fundamental theoretical basis of the observed photophysics.

The series of QD sizes incorporated into the assemblies facilitates comparison of the response as a function of the spectral overlap between the gold nanoparticle LSPR energy and emission of the QDs, as shown in Figure S2. There is a higher degree of spectral overlap between Q550 and the gold nanospheres (AuNS$_{10}$) relative to these with Q570, Q610 and Q650 ($4.78 > 3.89 > 2.78 >1.62 \times 10^{-12} \SI{}{cm}^{6}\SI{}{\milli\mole}^{-1}$) respectively. 

\subsubsection{Hybrid Dimers}
\begin{figure}[!ht]
    \includegraphics[width=\textwidth]{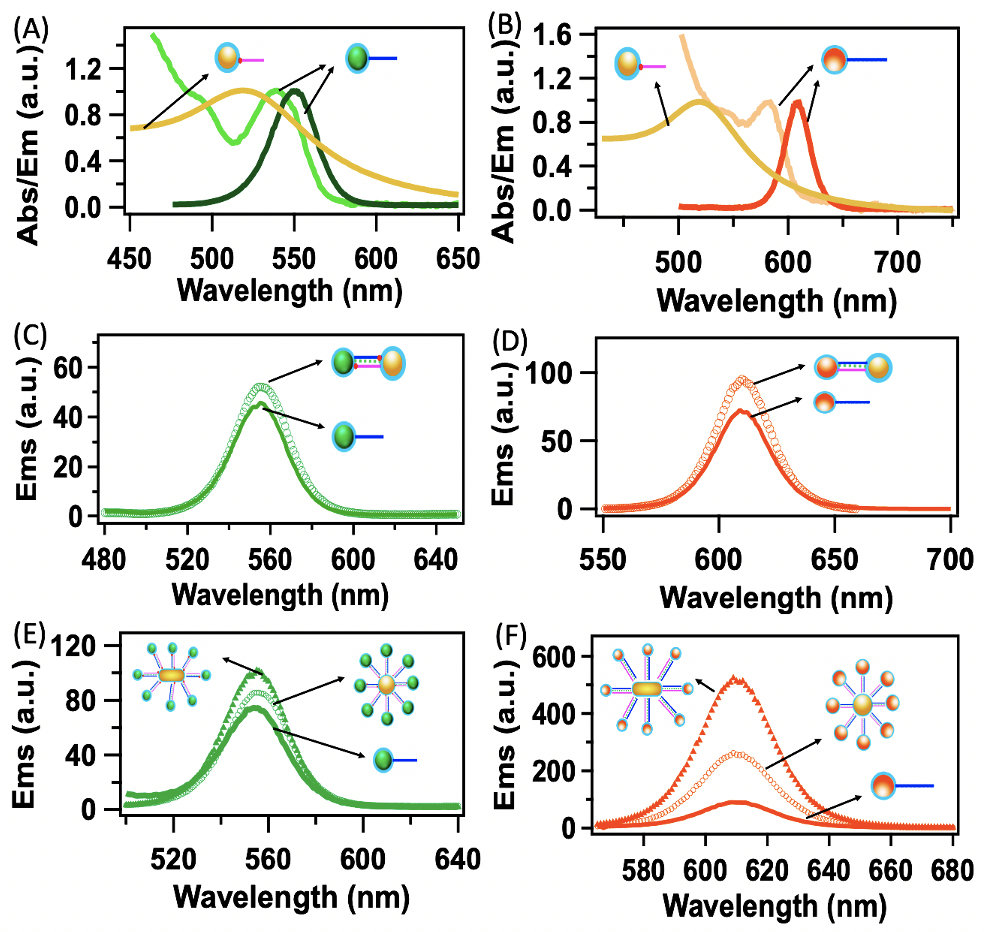}
\caption{Absorption and emission spectrum of QD overlapped with the extinction spectrum of metal nanoparticle. (A) Q550 (light and dark green) and AuNS$_{10}$ (brown) (B) Q610 (light and dark orange) and AuNS$_{10}$ (brown). Steady-state PL measurements of the hybrid assemblies developed by 30 bp DNA ($\sim\SI{10}{\nano\meter}$ interparticle separation) (C) Q550-Ref (solid line green) and Q550-AuNS$_{10}$ dimers (circle) at 2 $\mu$M concentration. (D) Q610-Ref (plain orange) and Q610-AuNS$_{10}$ dimers (circle orange) at at 2 $\mu$M concentration. E) Q550-Ref (plain green), Q550-AuNS$_{10}$ COSA (green circle)and Q550-AuNR$_{2.7}$ COSA (green triangle) at 3 $\mu$M QD concentration. The average number of satellites are  $9\pm2$ for Q550-AuNS$_{10}$ and $18\pm4$ for Q550-AuNR$_{2.7}$ COSA assemblies. (F) Q610-Ref (plain orange), Q610-AuNS$_{10}$ COSA (circle orange) and Q610-AuNR$_{2.7}$ COSA (triangle orange) assemblies at 3 $\mu$M QD concentration. The average number of satellites are $5\pm2$ for Q610-AuNS$_{10}$ and $13\pm4$ for Q610-AuNR$_{2.7}$ COSA assemblies.}\label{Fig:Steady_state_Q550-Q610}
\label{Steady_state_Q550-Q610}
\end{figure}

The emission spectra of the Q550-AuNS$_{10}$ dimers with 30 bp DNA are shown in Figure \ref{Steady_state_Q550-Q610}C. The photoluminescence intensity in the Q550-AuNS$_{10}$ dimer increases by 19\% compared to the monomer (Figure \ref{Steady_state_Q550-Q610}A) before hybridisation (24\% for the 100 bp dimer). The observed enhancement is a consequence of the balance between the enhancement of the electric field and quenching of the QD PL, both of which are distance-dependent, at 10 and 34 nm from the metal particle.\cite{kulakovich2002enhanced,ribeiro2013enhanced}

The PL decays were fit using a tri-exponential model, reflecting the broad distribution of emission rates of the QD, and are tabulated in Table \ref{Lifetimes} for dimers linked by 30 bp DNA linkers and Table S3 for 100 bp DNA linkers. The decay histogram for Q550-AuNS$_{10}$ (30 bp DNA linker) is shown in Figure \ref{TCSPC_Q550-Q610}A. The average lifetime of Q550 is reduced from that of the reference QD, $\SI{9.9}{\nano\second}$, to $\SI{7.8}{\nano\second}$ upon incorporation of the QD into the dimer (Table \ref{Lifetimes}), a reduction of $(\SI{7.8}{\nano\second}/\SI{9.9}{\nano\second})=0.79$. The increase in the steady-state emission intensity  (Figure \ref{Steady_state_Q550-Q610}C) and decrease in lifetime for Q550-AuNS$_{10}$ is consistent with literature reports,\cite{li2010enhancing,cohen2012exciton,song2005large,Nepal_2013_Acsnano} and attributed to coupling of the QD excited state to the AuNP LSPR.\cite{vaishnav_2018_Phy_Chem_C} The reduction in the degree of quenching of the lifetime of the 100 bp DNA-based Q550-AuNS$_{10}$ hybrid dimer $(\SI{7.1}{\nano\second}/\SI{8.7}{\nano\second})=0.82$ (Figure S6C, Table S3) compared with that of the shorter linker, along with the greater enhancement in the steady-state PL intensity for this larger separation between the QD and gold nanosphere, is also consistent with literature reports.\cite{li2010enhancing} 

\begin{table}
  \caption{Photophysics of the hybrid assemblies fabricated using 30 bp DNA linkers.}
  \label{Lifetimes}
  \begin{tabular}{llccccc}
    \hline
    Assembly & NP 1 & NP 2 & $\sharp$ NP 2 & $J$ & $\langle\tau\rangle$ ($\SI{}{\nano\second}$) & $\langle\tau\rangle$ ($\SI{}{\nano\second}$) \\
    Type & & & & ($\times10^{-12}$ $\SI{}{\centi\meter}^{6}$ & QD-ref & Assembly \\
    & & & & $\SI{}{\milli\mole}^{-1}$) & & \\
    \hline
    Dimer   & AuNS$_{10}$ & Q550 & 1 & 4.78 & 9.9 $\pm 0.7$  & 7.8 $\pm 0.4$ \\
    Dimer & AuNS$_{10}$ & Q610 & 1 & 2.78 & 12.9 $\pm 0.4$  & 16.3 $\pm 0.3$  \\
    COSA  & AuNS$_{10}$ & Q550 & 9 $\pm2$ & 4.78& 8.4  $\pm 0.4$  & 6.5 $\pm 0.4$   \\
    COSA  & AuNS$_{10}$ & Q570 & 7 $\pm2$ & 3.89& 13.1 $\pm0.4$  & 14.4 $\pm0.3$  \\
    COSA  & AuNS$_{10}$ & Q610 & 5 $\pm2$ & 2.78 & 12.0 $\pm0.3$  & 19.8 $\pm0.5$  \\
    COSA  & AuNS$_{10}$ & Q650 & 5 $\pm2$ & 1.62 & 9.0 $\pm0.4$  & 9.7 $\pm0.4$  \\
    COSA  & AuNR$_{2.7}$ & Q550 & 18 $\pm4$ & 1.61 & 8.4 $\pm0.4$  & 6.7 $\pm0.4$  \\
    COSA  & AuNR$_{2.7}$ & Q570 & 16 $\pm4$ & 2.80 & 13.1 $\pm0.4$  & 16.0 $\pm0.6$  \\
    COSA  & AuNR$_{2.7}$ & Q610 & 13 $\pm4$ & 2.88 & 12.0 $\pm0.3$  & 20.3 $\pm0.5$  \\
    COSA  & AuNR$_{2.7}$ & Q650 & 13 $\pm4$ & 3.93 & 9.0 $\pm0.4$  & 10.7 $\pm0.3$  \\
    \hline
\end{tabular}\\
\end{table}

In contrast to the results for the Q550 dimer, those for Q610-AuNS$_{10}$, within which the LSPR and QD emission have little spectral overlap (Figure \ref{Steady_state_Q550-Q610}B), show both an increase in both the steady-state PL intensity and the PL lifetime of the QD, that is, the photoluminescence lifetime of the QD is not quenched, but extended (a decrease in the emission rate). The steady-state photoluminescence intensity of Q610-AuNS$_{10}$ dimer linked with a 30 bp DNA linker increases by 23\% compared to the reference (Figure \ref{Steady_state_Q550-Q610}D), a slightly larger enhancement compared to that for Q550-AuNS$_{10}$. A lengthening of the QD lifetime from $\langle\tau\rangle = \SI{12.9}{\nano\second}$ to $\langle\tau\rangle = \SI{16.4}{\nano\second}$ upon incorporation of the QD into the dimer (30 bp linker length), equating to an increase of $(\SI{16.4}{\nano\second}/\SI{12.9}{\nano\second})=1.3$ is apparent (Figure \ref{TCSPC_Q550-Q610}B). For Q610-AuNS$_{10}$ assemblies linked via a 100 bp DNA linker, steady-state PL intensities increased by 13\% (Figure S6B), and the PL lifetime increased by 1.2 (Figure S6D) relative to the reference. Hence, also in contrast to those for the Q550-AuNS$_{10}$ dimer, the changes in PL intensity and lifetime are both more pronounced for the shorter interparticle separation. 

The reduction of lifetimes at one end of the spectral overlap scale (i.e. high spectral overlap), combined with the increase in the lifetime for low spectral overlap provides the opportunity for full control of the emission decay rate of the QD in metal-semiconductor hybrid systems. The level of control extends across a time period greater than previously appreciated as a consequence of the increase in the lifetime when the spectral overlap between the nanoparticles is detuned. For all systems, irrespective of the change in emission rate, an increase in the steady-state photoluminescence intensity was observed, although the relative trends with respect to their interparticle separation are different due to the competing factors. 

\subsubsection{Hybrid Core-satellite Assemblies}
The core-satellite assemblies contain multiple QD particles around a metal nanoparticle, correlating with many donor/single acceptor systems in the framework of energy transfer systems. Relatively little change in the degree of enhancement of the steady-state PL intensity (increase by 29\%, Figure \ref{Steady_state_Q550-Q610}E)) or lifetime quenching (0.79, Table \ref{Lifetimes}) is observed upon increasing the number of QDs around Q550-AuNS$_{10}$-COSA with 30 bp linking DNA (Figure \ref{COSA_gels}C) compared to the dimer.

Hybrid Q610-AuNS$_{10}$ COSA assemblies with 30 bp linker length, however, have a greater degree of enhancement in the PL intensity (48\% enhancement), shown in Figure \ref{Steady_state_Q550-Q610}F, compared to the dimer assembly. Similarly, multiple Q610 quantum dots around a single gold nanoparticle also results in a large increase, ($\SI{19.8}{\nano\second}$/$\SI{12.0}{\nano\second})=1.65$, in the PL lifetime relative to the monomer (Figure \ref{Steady_state_Q550-Q610}F, Table \ref{Lifetimes}), and 1.2 relative to the dimer. Enhancement in PL intensity and lengthening of the average lifetime for hybrid assemblies incorporating Q570 and Q650, which have varying degrees of de-tuning of their emission with the LSPR of AuNS$_{10}$ and 30 bp linkers, is also observed (Figure S9). The degree of spectral overlap for the Q650 hybrid assembly is lower than Q610. A lower enhancement in the magnitude of PL intensity and average lifetime for the Q650 than Q610 hybrid assemblies is however observed. This is attributed to an increase in the center to center spacing between the AuNP and Q650 owing to the bigger size of Q650, with potentially some effect from the low quantum yield of Q650. The quenching of the quantum emitter lifetime is more pronounced for the QD with a higher spectral overlap with AuNS$_{10}$, Q550-AuNS$_{10}$-COSA, for which the quenching dominates the optical response at short interparticle separations. For all other assemblies, a trade-off exists between the degree of spectral overlap and interparticle separation. 

Consistent with the results for dimers, increasing the interparticle separation within the core-satellite structures based on AuNS$_{10}$ led to an increase in the steady-state photoluminescence intensity to 31\% and 35\% for Q550-AuNS$_{10}$-COSA and Q610-AuNS$_{10}$-COSA assemblies, respectively. The lifetime decreased by $\sim\SI{1}{\nano\second}$ for Q550-AuNS$_{10}$-COSA and increased by $\sim\SI{1}{\nano\second}$ for Q610-AuNS$_{10}$-COSA assemblies (Figure S7, Table S4). Control of the assembly structure thus allows tuning of the degree of photoluminescence enhancement and emission rate.

\begin{figure}[!ht]
    \includegraphics[width=\textwidth]{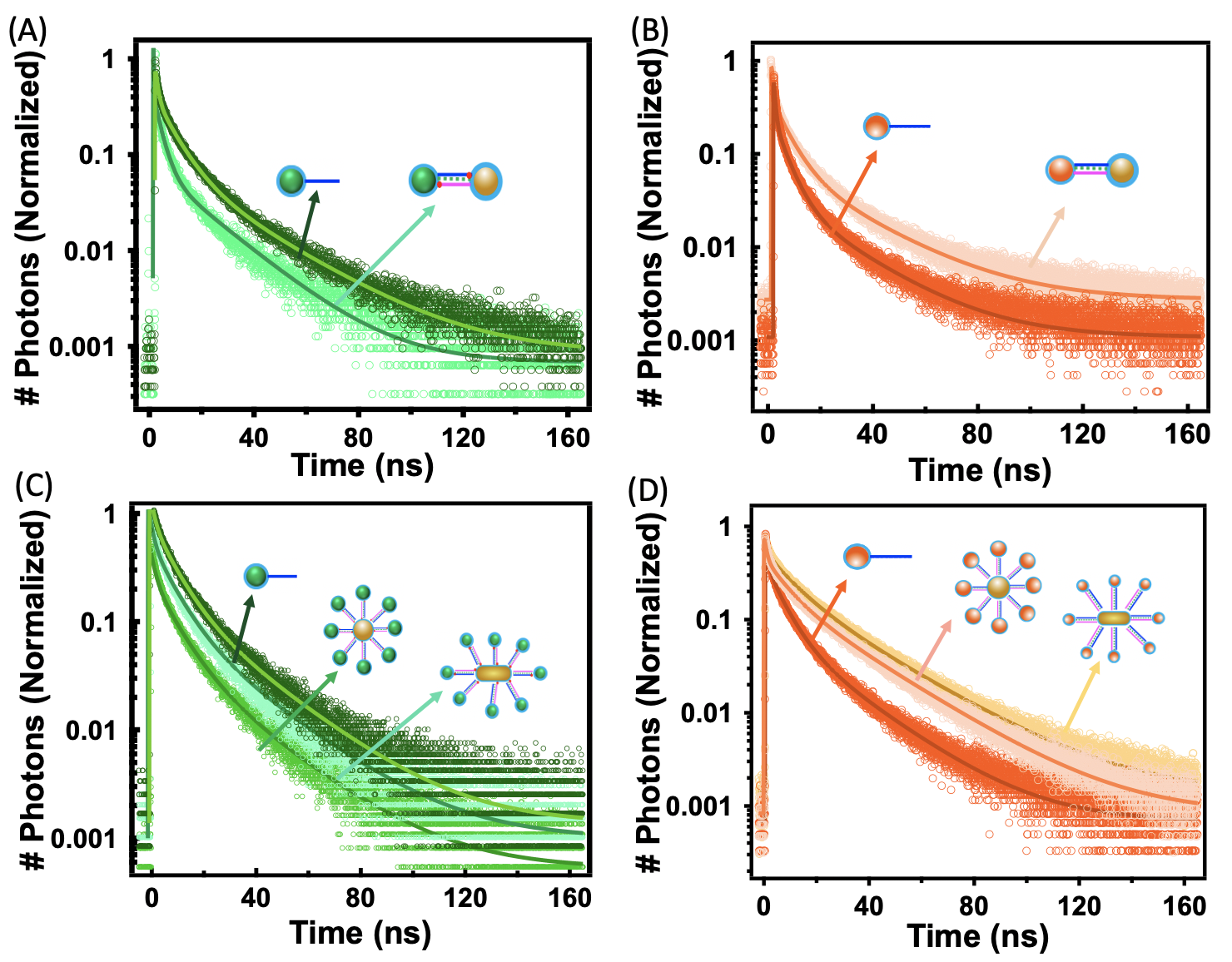}
\caption{Normalized fluorescence lifetime decays of an ensemble of hybrid assemblies with 30 base pair DNA ($\sim\SI{10}{\nano\meter}$ interparicle separation) at $\SI{425}{\nano\meter}$ excitation. (A) Q550-Ref (dark green) and Q550-AuNS$_{10}$ dimers (light green) (B) Q610-Ref (dark orange) and Q610-AuNS$_{10}$ dimers (light orange) (C) Q550-Ref (dark green), Q550-AuNS$_{10}$ (medium green) and Q550-AuNR$_{2.7}$ (light green) COSA assemblies (D) Q610-Ref (dark orange) and Q610-AuNS$_{10}$ (light orange) and Q610-AuNR$_{2.7}$ (medium orange) COSA.}\label{Fig:Exp_decay}
\label{TCSPC_Q550-Q610}
\end{figure}

The enhancement effect in the PL intensity and the change in the lifetime values in the hybrid assemblies are more pronounced upon the incorporation of larger, asymmetric plasmonic particles such as gold nanorods. For nanorod assemblies with 30 bp DNA linkers, the PL intensity increased from the reference PL values to 45\% for Q550-AuNR$_{2.7}$-COSA assemblies and 75\% for Q610-AuNR$_{2.7}$-COSA assemblies (Figure \ref{Steady_state_Q550-Q610}E and F). The lifetime of Q550 decreased by ($\SI{6.73}{\nano\second}$/$\SI{8.44}{\nano\second})=0.80$ in Q550-AuNR$_{2.7}$-COSA assemblies and increased by ($\SI{20.4}{\nano\second}$/$\SI{12.0}{\nano\second})=1.6$ for Q610-AuNR$_{2.7}$-COSA assemblies (Figure \ref{TCSPC_Q550-Q610}C and D). With 100 bp DNA linkers, the same assemblies showed an increase in the intensity of the PL, although not to the same degree as observed for the shorter linkers, with the lifetime for the Q550-AuNR$_{2.7}$-COSA assembly showing a similar decrease and a smaller increase in the lifetime for Q610-AuNR$_{2.7}$-COSA (Figure S7). The photophysical properties of the AuNR$_{2.7}$-COSA are thus consistent with both AuNS$_{10}$-COSA and hybrid dimers assemblies, although of a greater magnitude. This is presumably due to the strong localisation of the near-field of the nanorod at the tips. 

Previous literature shows that the local refractive index of the metal nanoparticle has a large impact on the decay kinetics of QDs in metal-semiconductor assemblies \cite{colas2012mie}. In the next section, we model the optical properties of these systems to fully understand their origin. 

\subsection*{Theoretical insights}

\subsubsection*{Impact of the local electromagnetic environment on spontaneous emission} 

The transition rate (or the transition probability per unit time) of a quantum emitter from an excited state $| e\rangle$ to a ground state $| g\rangle$ follows the Fermi's golden rule given by \cite{fox2006quantum},
\begin{equation}\label{Eq:Fermi_Golden_rule}
\gamma_{e\rightarrow g} = \frac{2\pi}{\hbar^2}|M_{eg}|^2 g(\omega).
\end{equation}
The density of states $g(\omega)$ of the electromagnetic environment of the emitter is defined such that $g(\omega) d\omega$ is the number of photon states per unit volume that fall within the angular frequency range $\omega$ to $\omega+d\omega$ \cite{fox2006quantum, premaratne2021theoretical}. The transition matrix element $M_{eg}$ is obtained as,
\begin{equation}\label{Eq:M_eg}
M_{eg} = \langle e|\hat{H}'|g \rangle = -\bm{\mu}_{eg} . \bm{\mathcal{E}},
\end{equation}
where $\hat{H}'$ is the perturbation Hamiltonian caused by light, $\bm{\mu}_{eg}$ is the electric dipole moment of the transition and $\bm{\mathcal{E}}$ is the electric field amplitude experienced by the quantum emitter. It is evident that the transition rate is determined by the interplay between the electric field experienced by the emitter and the density of states. 
	
\subsubsection*{Damped nanoscale cavity analogy for gold nanoparticles} 

Let us first analyse how an MNP placed in nanoscale proximity alters the electromagnetic environment of a coupled QD using a cavity analogy. The total cavity decay rate $\Gamma_n$ (should not be confused with the emitter decay rate $\gamma$), the quality factor $Q_n$ and the effective mode volume $V_n$ of the $n^\text{th}$ cavity-like plasmonic mode of a spherical metal nanoparticle with radius $r_\text{m}$ can be estimated using the Drude approximation as $\Gamma_n = \Gamma_\text{abs} + \Gamma_n^\text{rad}$ and $Q_n = \omega_n/\Gamma_n$ where \cite{colas2012mie},  

\begin{align}
\Gamma_n^\text{rad} &= \frac{(2n+1)\epsilon_\text{b}}{n\epsilon_\infty+(n+1)\epsilon_\text{b}}\omega_n\frac{(n+1)(k_\text{b} r_\text{m})^{2n+1}}{n(2n-1)!!(2n+1)!!},\\
\omega_n &= \omega_p\sqrt{n\big/\left[n\epsilon_\infty+(n+1)\epsilon_\text{b}\right]}\\
V_n &= \frac{n\epsilon_\infty+(n+1)\epsilon_\text{b}}{(2n+1)\epsilon_\text{b}}\frac{9}{(2n+1)(n+1)}\frac{4}{3}\pi r_\text{m}^3.
\end{align}
In the above equations, $\omega_p$ and $\Gamma_\text{abs}$ represent the bulk plasma frequency and the ohmic loss rate of the metal, $\Gamma_n^\text{rad}$ is the radiative scattering rate, $\epsilon_\text{b}$ is the relative permittivity of the non-magnetic submerging medium (where $n_\text{b} = \sqrt{\epsilon_\text{b}}$ is the background refractive index), $\epsilon_\infty$ is the contribution of bound electrons in the metal dielectric constant and $k_\text{b} = n_\text{b} k = 2\pi n_\text{b}/\lambda$ is the wavenumber in the background medium for free space wavelength $\lambda$ (and free space wavenumber $k$). 

\begin{figure}[t!]  
	\includegraphics[width=\columnwidth]{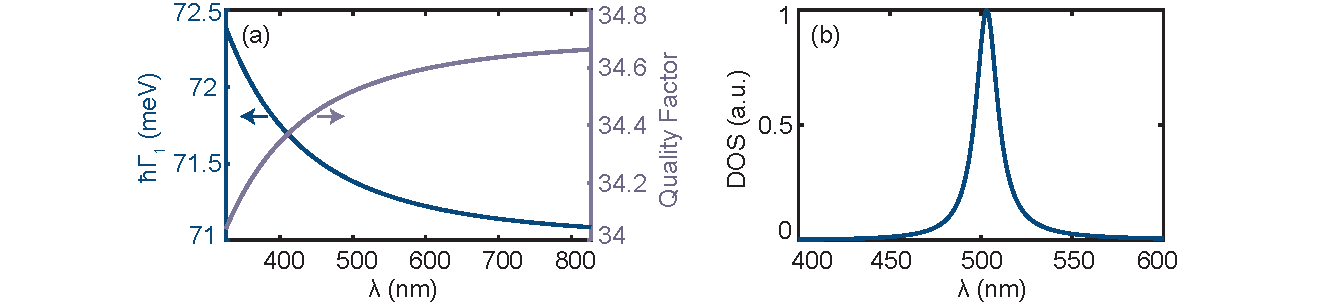}
	\centering
	\caption{(a) Estimated cavity decay rate, quality factor, and (b) qualitative variation of the density of states for the dipolar ($n=1$) mode of an $r_\text{m}\sim\SI{5}{\nm}$ gold nanoparticle submerged in water. \label{Fig:Gamma_and_Q_fac}}
\end{figure}

We estimate a dipole ($n=1$) mode volume $V_1 \sim \SI{1970}{\nano\meter\cubed}$, and the plasmonic decay rate and quality factor variations shown in Figure\ \ref{Fig:Gamma_and_Q_fac}(a), for a gold nanoparticle of radius $r_\text{m}\sim\SI{5}{\nano\meter}$ submerged in water, using the above equations and the following parameters from literature: $\hbar\omega_p \approx \SI{9.02}{\electronvolt}$, $\hbar\Gamma_\text{abs} \approx \SI{0.071}{\electronvolt}$ \cite{raza2015nonlocal}, $\epsilon_\text{b}\approx1.78$ and $\epsilon_\infty \approx 9.84$ \cite{kolwas2010plasmonic}.

It has been shown that rather low-quality factors ranging from about 10 to 100 such as those in Figure\ \ref{Fig:Gamma_and_Q_fac}(a), accompanied by small effective volumes in the nanometer scale result in high-quality factor/effective volume ratios that indicate efficient coupling to quantum emitters generally \cite{colas2012mie}. This is in contrast to conventional diffraction-limited micro-cavities with comparatively much higher volumes that efficiently couple to emitters thanks to their very high-quality factors \cite{colas2012mie}. Thus, our gold nanoparticle submerged in water is expected to behave as a damped, low-quality factor nanoscale cavity capable of efficiently coupling to emitters, while altering their electromagnetic environments with reasonably high bandwidths. 

Using the aforementioned cavity analogy, we can now estimate the density of states $g(\omega)$ of the MNP cavity for a given single mode with half-width $\Delta\omega_c$ and cavity resonance $\omega_c$ as: $g(\omega) \sim (2\Delta\omega_c)\big/\lbrace \pi [4(\omega-\omega_c)^2 + \Delta\omega_c^2]\rbrace$ \cite{fox2006quantum}. The qualitatively estimated density of states variation for the dipole mode of the earlier MNP obtained using $\Delta\omega_c \sim \Gamma_1$ and $\omega_c \sim \omega_1$ is depicted in Figure\ \ref{Fig:Gamma_and_Q_fac}(b). This results in a Lorentzian-shaped density of states variation suggesting that an MNP seems more likely to accept a photon emitted by a nearly resonant QD due to the high density of states, whereas this ability depletes as the QD becomes off-resonant. In essence, we can interpret the experimentally observed QD decay rate modification as a manifestation of the cavity nature of gold nanoparticles, where the emission is enhanced when the \enquote{cavity} is tuned (in Figure\ \ref{Fig:Exp_decay}(A) for Q550-AuNS$_{10}$) and the emission is suppressed when it is detuned (in Figure\ \ref{Fig:Exp_decay}(B) for Q610-AuNS$_{10}$), as suggested by equation (\ref{Eq:Fermi_Golden_rule}).

In conventional resonant low Q cavities, the rapidity of emitter decay has been reported to increase with the number of emitters \cite{seke1987n}. This behavior is marginally mimicked by Q550-AuNS$_{10}$ core-satellite assemblies with average normalized lifetime $\SI{6.5}{\nano\second}/\SI{8.4}{\nano\second} \approx 0.77$ (see Table \ref{Lifetimes}), in comparison to Q550-AuNS$_{10}$ dimers with average normalized  lifetime $\SI{7.8}{\nano\second}/\SI{9.9}{\nano\second} \approx 0.79$. Note that the observed difference between the dimer and core-satellite assemblies stems not just from the number of surrounding emitters but also from the (unquantified) increase in the local refractive index of the MNP due to the larger number of DNA strands in the latter case.  It has also been reported in the context of conventional cavities that collective emission effects can be appreciably suppressed by increasing the detuning of the cavity from the emitter resonance for multiple emitter assemblies \cite{seke1987n}. This is qualitatively mimicked by our experimental observations for Q610-AuNS$_{10}$ Dimer (average normalized lifetime $\SI{16.3}{\nano\second}/\SI{12.9}{\nano\second} \approx 1.26$) and core-satellite assemblies (average normalized lifetime $\SI{19.8}{\nano\second}/\SI{12}{\nano\second} \approx 1.65$). While reporting these qualitative similarities, we also emphasize that MNPs are not expected to ideally mimic conventional, diffraction-limited linear optical cavities. MNPs form feedback dipoles in response to all proximal dipoles \cite{artuso2012optical, hapuarachchi2020influence}, which causes emitter-AuNP assemblies to behave as non-linear systems.

\subsubsection*{MNP-QD dimers as semi-classical open quantum systems}
Adopting an alternative viewpoint where we can account for the aforementioned non-linearities, we now model the MNP using the generalized nonlocal optical response theory (GNOR) \cite{raza2015nonlocal} and the QD excitons as open quantum systems.

\begin{figure}[t!]  
	\includegraphics[width=\columnwidth]{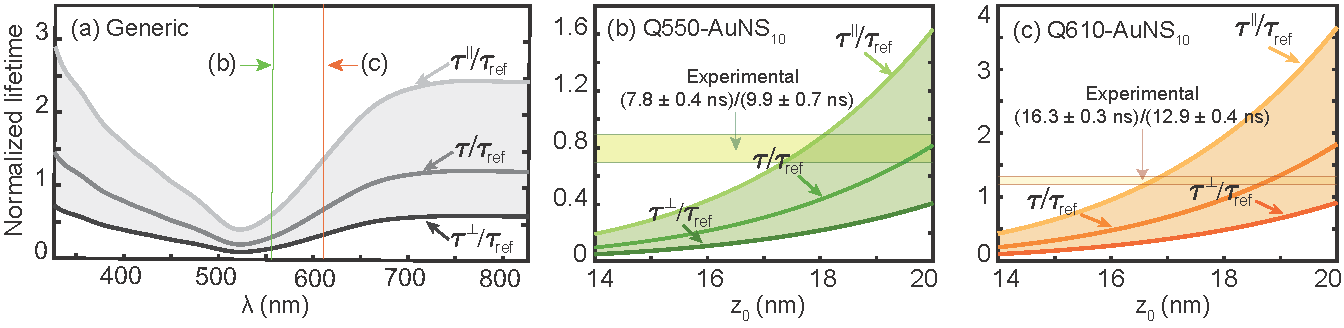}
	\centering
	\caption{(a) Estimated lifetimes for a generic emitter near a metal nanoparticle calculated using equations (\ref{Eq:gamma_perp_NL_main}), (\ref{Eq:gamma_parallel_N_mainL}) and (\ref{Eq:gamma_rand_main}). AuNS parameters used to obtain the GNOR based polarizability are same as those in Figure\ \ref{Fig:Dimer_ss_and_temporal}. Sub-figures (b) and (c) depict cross sections of (a) plotted against the inter-particle centre separation $z_0$ at $\lambda = \SI{556}{\nano\meter}$ (emission maximum of Q550) and $\lambda =\SI{610}{\nano\meter}$ (emission maximum of Q610), respectively. \label{Fig:normalized_decay_and_lifetimes}} 
\end{figure}

Incorporating GNOR based non-locality corrections to the dipolar mode of the Mie expansion, the normalized decay rate for a generic dipolar emitter placed at a small centre separation $z_0$ from a spherical metal nanoparticle can be estimated as, \begin{subequations}
	\begin{align}
	\frac{\gamma^\perp}{n_\text{b}\gamma_0} = \frac{\tau_\text{ref}}{\tau^\perp} &\approx \frac{6  \Im\left[\alpha_1^\text{NL}\right]}{k_\text{b}^3 z_0^6}, \label{Eq:gamma_perp_NL_main}\\
	\frac{\gamma^\parallel}{n_\text{b}\gamma_0}  = \frac{\tau_\text{ref}}{\tau^\parallel} &\approx \frac{3 \Im\left[\alpha_1^\text{NL}\right]}{2 k_\text{b}^3 z_0^6}, \label{Eq:gamma_parallel_N_mainL}
	\end{align}
\end{subequations}
where $\gamma^\perp$, $\gamma^\parallel$ and $\gamma_0$ denote the spontaneous emission rates of dipole emitters oriented perpendicularly and parallelly to the MNP surface, and free-space spontaneous decay rate, respectively. The decay rate of the isolated emitter in a medium of refractive index $n_\text{b}$ is obtainable as $\gamma_\text{ref} = n_\text{b}\gamma_0$ \cite{duan2006dependence}, and $\tau$ (with each respective subscript/superscript) denotes the emitter excitation lifetime. See the supplementary material for details, including the full form of the GNOR based MNP dipolar polarizability $\alpha_1^\text{NL}$. The decay rate $\gamma$ of a randomly oriented emitter in a hybrid dimer can be estimated as \cite{colas2012mie},
\begin{equation}\label{Eq:gamma_rand_main}
\gamma = (\gamma^\perp + 2\gamma^\parallel)\big/3.  
\end{equation}
Figure\ \ref{Fig:normalized_decay_and_lifetimes}(a) shows that for a generic emitter, the normalized excitation lifetime is expected to increase as the emission frequency moves away from the MNP resonance (and vice versa) as per equations (\ref{Eq:gamma_perp_NL_main}) and (\ref{Eq:gamma_parallel_N_mainL}), in line with our earlier observations and claims. It can also be seen from Figure\ \ref{Fig:normalized_decay_and_lifetimes}(b) and (c) that the PL lifetimes of both Q550 and Q610 are theoretically expected to decrease in response to a gold nanoparticle approaching closer (decreasing $z_0$), in line with the previous single dimer based observations in literature \cite{Ratchford_2011_Nanoletters}. The lifetimes of QDs we reported experimentally (for dimers in Table \ref{Lifetimes}) are averages resulting from the cohort of dimers present in colloidal samples, where the centre separations may vary around $\sim \SI{17}{\nano\meter}$.

\begin{figure}[t!]  
	\includegraphics[width=\columnwidth]{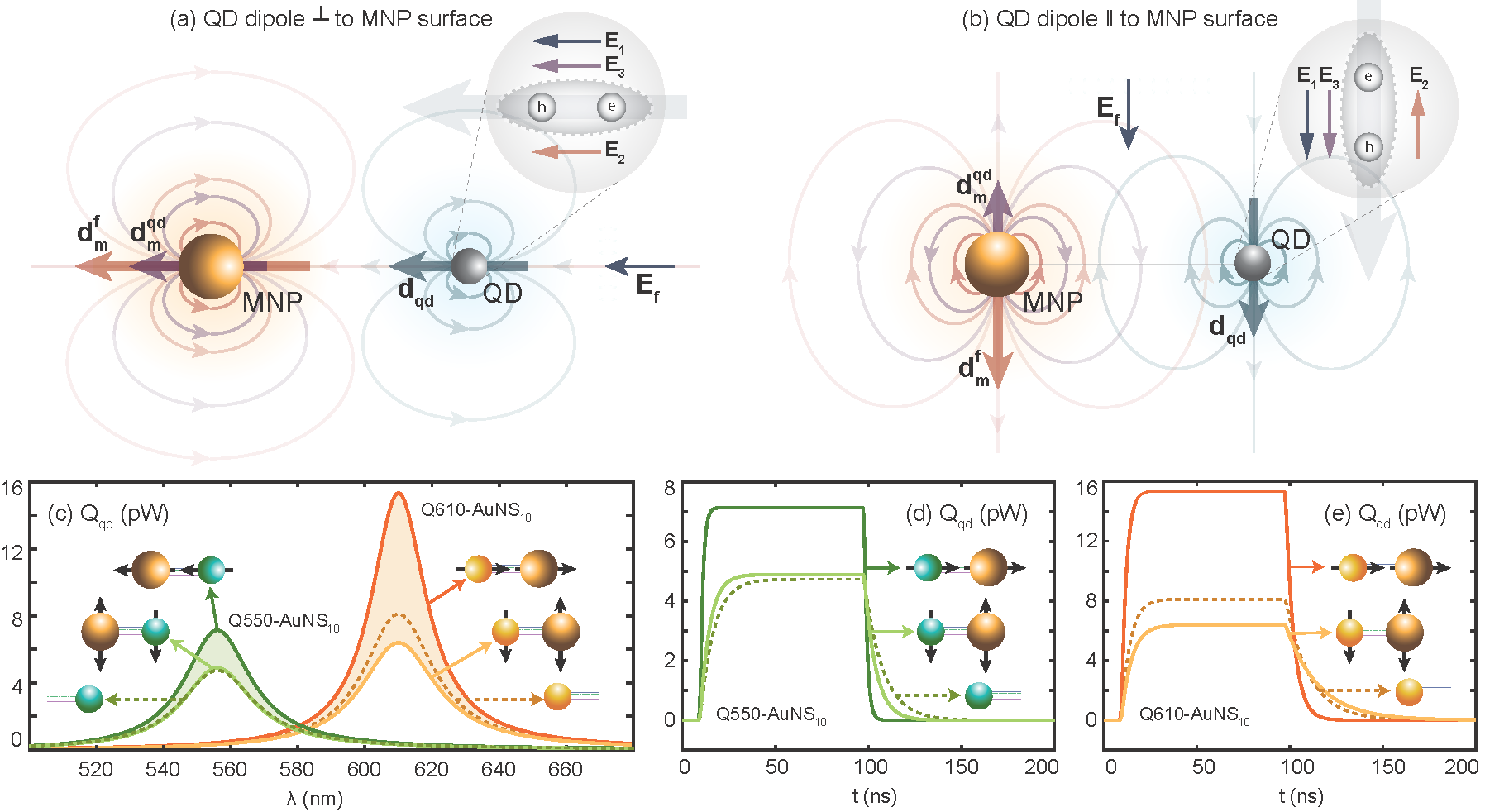}
	\centering
	\caption{(a) and (b) depict schematic diagrams for the cases where the QD dipoles are perpendicular and parallel to the MNP surface. The exciton formed in the QD experiences three external fields: $\bm{E_1}$ (external field $\bm{E}_\text{\textbf{f}}$ screened by the QD material), $\bm{E_2}$ (field emanated by the dipole $\bm{d}_\text{\textbf{m}}^\text{\textbf{f}}$ formed in the MNP in response to $\bm{E}_\text{\textbf{f}}$), and $\bm{E_3}$ (field emanated by the MNP dipole $\bm{d}_\text{\textbf{m}}^\text{\textbf{qd}}$ formed in response to the QD dipole $\bm{d}_\text{\textbf{qd}}$). Theoretically generated steady-state and temporal emission spectra for the extreme orientations of Q550-AuNS$_{10}$ and Q610-AuNS$_{10}$ (with the isolated QD plots as references) are shown in sub-figures (c), (d) and (e). The shaded regions depict where the emission from intermediate orientations would fall. Theoretical formalism and all  parameters used are detailed in the supplementary material. \label{Fig:Dimer_ss_and_temporal}}
\end{figure}

We model the exciton that forms in the QD as a coherently illuminated two-level-atom (TLA) coupled to the plasmonic field of the MNP at one of the two extreme orientations (QD transition dipole parallel or perpendicular to the MNP surface as portrayed in Figure\ \ref{Fig:Dimer_ss_and_temporal}). The TLA (both in isolation and in the presence of the MNP) is assumed to behave as an open quantum system that undergoes Markovian interactions with the submerging environment. This TLA  model, where we equate the excitonic energy gap to the experimentally observed emission spectral peak of a given QD is only expected to approximate the emission behaviour, which is the focus of our study. It is not expected to sufficiently capture the phonon-assisted absorption mechanism of the QD which results in an absorption peak that does not overlap with the emission peak (unlike for an ideal TLA). Following the detailed formalism summarized in the supplementary material, we arrive at the following optical Bloch equations for the density matrix elements of the TLA, under the influence of an external electric field and the MNP,

\begin{subequations} \label{Eq:Bloch_equations_main}
	\begin{align}
	\dot{\rho}_\text{ee} &= -\frac{\rho_\text{ee}}{\tau^\measuredangle} + i\Omega^r\tilde{\rho}_\text{ge} - i\Omega^{r*}\tilde{\rho}_\text{eg} \label{Eq:rhoee_tilde_dot_main},\\
	\dot{\rho}_\text{gg} &= \frac{\rho_\text{ee}}{\tau^\measuredangle} - i\Omega^r\tilde{\rho}_\text{ge} + i\Omega^{r*}\tilde{\rho}_\text{eg} \label{Eq:rhogg_tilde_dot_main},\\
	\dot{\tilde{\rho}}_\text{eg} &=-\left[ i(\omega_0 - \omega) + 1\big/T^\measuredangle\right]\tilde{\rho}_\text{eg} + i\Omega^r(\rho_\text{gg} - \rho_\text{ee}), \label{Eq:rhoeg_tilde_dot_main}
	\end{align}
\end{subequations}
where $\rho_{ij}$ is the $ij^\text{th}$ density matrix element of the TLA, considering the basis states $\lbrace|g\rangle, |e\rangle\rbrace$,  $\rho_\text{eg} = \tilde{\rho}_\text{eg}e^{-i\omega t}$, $\tilde{\rho}_\text{ge} = \tilde{\rho}_\text{eg}^*$, $\tau^\measuredangle$ ($\measuredangle=\lbrace\perp, \parallel\rbrace$) is the corresponding emitter decay time (lifetime), $T^\measuredangle$ is the corresponding dephasing time, $\omega_0$  is the emitter resonance and $\omega$ is the angular frequency of the incoming radiation. The effective Rabi frequency in the presence of the MNP ($\Omega^r$) is related to the slowly varying positive frequency amplitude ($\tilde{E}^+_\text{qd}$) of the total field experienced by the QD exciton ($E_\text{qd}$) via the QD dipole moment element $\mu$ as $\tilde{E}^+_\text{qd} = \hbar\Omega^r/\mu$. ${E}_\text{qd}$ is the sum of three field components incident on the QD exciton: $E_1$ (external field $E_f$, screened by the QD material), $E_2$ (field emanated by the dipole formed in the MNP in response to $E_f$), and $E_3$ (field emanated by the dipole formed in the MNP in response to the QD dipole). A detailed discussion on dipole formation in nanohybrids can be found in reference \cite{hapuarachchi2020influence}. The total power emitted by the TLA takes the form $Q_\text{qd} \sim \hbar\omega_0 \rho_\text{ee}/\tau^\measuredangle$ for the extreme dimer orientations \cite{wrigge2008coherent, artuso2012optical}. For the isolated QD, $Q_\text{qd} \sim \hbar\omega_0 [\rho_\text{ee}]_\text{ref}/\tau_\text{ref}$, where $[\rho_\text{ee}]_\text{ref}$ denotes the isolated QD excited state population (see supplementary material for details). 

We used the above TLA-based open quantum system formalism to theoretically predict the behaviour of Q550-AuNS$_{10}$ and Q610-AuNS$_{10}$ dimers in extreme orientations. We assume that QDs are isotropic and form dipoles along the external field. We can see from Figure\ \ref{Fig:Dimer_ss_and_temporal}(a) that all fields experienced by a QD exciton constructively add in the same direction when the QD dipole is $\perp$ to the MNP surface (where all fields form along the axis of the hybrid), always resulting in a stronger field compared to the isolated QD (reference). This results in larger emission from the $\perp$ case compared to the reference for both Q550-AuNS$_{10}$ and Q610-AuNS$_{10}$ as observed in Figure\ \ref{Fig:Dimer_ss_and_temporal}(c) (where the darker solid lines for both dimers are larger than the reference). However, in the $\parallel$ orientation (see Figure\ \ref{Fig:Dimer_ss_and_temporal}(b) for the schematic), there can be constructive and destructive interference effects between the field components incident on the QD exciton, resulting in the emission of the dimer being enhanced or suppressed compared to the isolated QD depending on parameter variations. It can be observed from Figure\ \ref{Fig:Dimer_ss_and_temporal}(c) that Q610-AuNS$_{10}$ shows a large steady-state emission enhancement in the $\perp$ (axial) orientation
compared to Q550-AuNS$_{10}$. This suggests that the larger steady-state emission enhancement we experimentally observed for the Q610-AuNS$_{10}$ colloidal ensemble in Figure\ \ref{Fig:Steady_state_Q550-Q610}(D), compared to Q550-AuNS$_{10}$ in Figure\ \ref{Fig:Steady_state_Q550-Q610}(C), is likely to be attributable to the dominant contribution of axial field enhancement.

Temporal PL decay plots theoretically obtained for (single molecules of) both  Q550-AuNS$_{10}$ and Q610-AuNS$_{10}$ dimers are depicted in Figure \ref{Fig:Dimer_ss_and_temporal}(d) and (e). We only focus on the emission phase of these TLA-based time-domain plots, which occurs after the electric field input is switched off near $\SI{100}{\nano\second}$. Comparison of Q610's steady-state isolated and dimer spectra in Figure\ \ref{Fig:Dimer_ss_and_temporal}(c) to its temporal plots in (e) reveals that the plasmonically enhanced $\perp$ case in the former subplot maps to an enhanced temporal decay while the plasmonically suppressed $\parallel$ case maps to a suppressed temporal decay (or an enhanced lifetime, as is also observable from Figure\ \ref{Fig:normalized_decay_and_lifetimes}). The larger decay rate observed for the $\perp$ orientation is expected to be caused by the larger electric field experienced by the QD in the $\perp$ orientation compared to the $\parallel$ case as suggested by the Fermi's golden rule in (\ref{Eq:Fermi_Golden_rule}). Thus, the suppressed QD decay rate experimentally observed for the colloidal Q610-AuNS$_{10}$ dimer ensemble in Figure\ \ref{Fig:Exp_decay}(B) is likely to be attributable to the dominant contribution of the $\parallel$ components, as suggested by equation (\ref{Eq:gamma_rand_main}). 

\subsection{Tunability of Surface Plasmon Resonance}

\begin{figure}[!ht]
    \includegraphics[width=\textwidth]{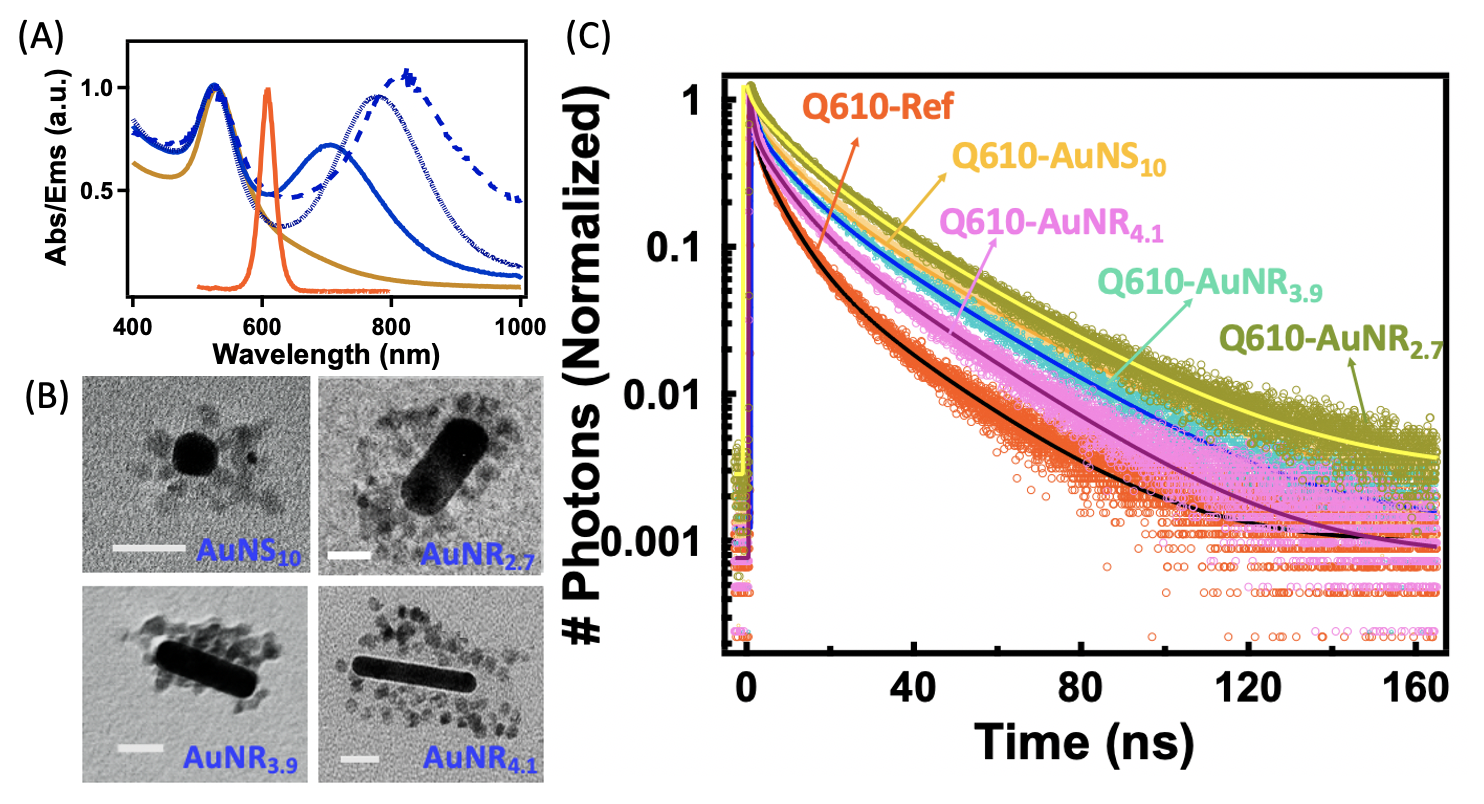}
\caption{(A) Spectral overlap of Q610 emission (plain orange) with respect to to the LSPR of AuNS$_{10}$ (brown) and variable aspect ratio AuNRs; AuNR$_{2.7}$ (plain blue), Q610-AuNR$_{3.9}$ (dotted blue) and Q610-AuNR$_{4.1}$ (dashed blue) (B) TEM characterization of the hybrid assemblies (C) Lifetime decays of Q610-Ref (dark orange), and hybrid assemblies Q610-AuNS$_{10}$ (light orange), Q610-AuNR$_{2.7}$ (dirty green) Q610-AuNR$_{3.9}$ (cyan), Q610-AuNR$_{4.1}$ (pink) with 30 bp DNA ($\sim$ 10 nm interparicle separation) at an excitation wavelength of 425 nm.}
\label{spectral_tuning}
\end{figure}

To obtain a complete picture of the QD-AuNP interaction, we investigated the coupling in hybrid assemblies with fixed QD emission and variable localized surface plasmon of the AuNP. For this purpose, hybrid assemblies of Q610 with AuNS$_{10}$ and different aspect ratios of gold nanorods i.e. 2.7, 3.9, and 4.1 were synthesised. The time-resolved measurements in Figure \ref{spectral_tuning} compare the emission lifetime of Q610-Ref and Q610-AuNR COSA assemblies with the different aspect ratios of the metal nanorod (fit parameters in Table S5). While a lengthening of the photoluminescence lifetime for all metal hybrid systems relative to the QD alone was observed, the emission rate of Q610-AuNR$_{2.7}$ (20.3 ns) is lower compared to that of Q610-AuNR$_{3.9}$ and Q610-AuNR$_{4.1}$ (Table \ref{Lifetimes for spectral tunability}). The spectral overlaps for these systems are not significantly different (Table \ref{Lifetimes for spectral tunability}). Nanorods are well known to have a larger near field enhancement at the tips of the nanorod, due to the longitudinal LSPR mode.\cite{xue2017tuning} Of the nanorods here, AuNR$_{2.7}$ has the greatest width (AuNR$_{2.7}$ = 20 nm, AuNR$_{3.9}$ = 10 nm, AuNR$_{4.1}$ = 7 nm). The wider the rod, the more QDs able to attach the tip, as apparent in the TEM images in Figure \ref{spectral_tuning} and also the higher probability that QDs will be present at the nanorod tip, that is, in the position of maximum field enhancement. Consequently, the greater lengthening of the average lifetime of Q610-AuNR$_{2.7}$ is attributed to the nanorod width. 

\begin{table}
  \caption{Photophysics of the hybrid assemblies with 30 bp DNA linkers with nanorods of different aspect ratios.}
  \label{Lifetimes for spectral tunability}
  \begin{tabular}{llcccc}
    \hline
    Sample  & NP 1  & NP 2  & $\sharp$ NP 2 & $J$ & $\langle\tau\rangle$ \\ 
    & & & & ($\times10^{-12}$ $\SI{}{\centi\meter}^{6}$ & \SI{}{\nano\second} \\
    & & & & $\SI{}{\milli\mole}^{-1}$) & \\
    \hline
    Q610-Ref  & Q610 & - & - & -  & 11.8  $\pm 0.4$ \\
    COSA  & AuNS$_{10}$ & Q610 & 5 $\pm2$ & 2.78  & 19.8  $\pm 0.5$ \\
    COSA  & AuNR$_{2.7}$ & Q610 & 13 $\pm4$ & 2.88& 20.3 $\pm0.3$ \\
    COSA  & AuNR$_{3.9}$ & Q610 & 11 $\pm4$ & 2.56 & 18.8 $\pm0.4$ \\
    COSA  & AuNR$_{4.1}$ & Q610 & 10 $\pm4$ & 2.11 &15.8  $\pm0.6$ \\
    \hline
\end{tabular}\\
\end{table}

\clearpage
\section{Conclusion}
We combine DNA-based self-assembly methods, optical spectroscopy, and theory to determine the fundamental interactions between plasmonic nanoparticles and quantum dots. Discrete assemblies incorporating quantum dots and gold nanoparticles are synthesised by translation of DNA-based assembly methods to CdSe QDs. The isolation of QDs containing only one strand of DNA per particle allows the formation of QD:AuNP dimers and core-satellite structures in high purity. The tight control of the nanoparticle sizes, stoichiometry, and interparticle separation (via DNA linker length) in the assemblies allows the interrogation of these precise structures using ensemble optical spectroscopy and theory.  

A gold nanoparticle immersed in water behaves as a nanoscale damped optical cavity that mimics the qualitative behaviour of conventional diffraction-limited optical cavities when modifying the emission characteristics of quantum emitters. We show experimentally and theoretically that the detuning between a metal nanoparticle and emitter can be effectively used for the regulation (enhancement/suppression) of spontaneous emission from quantum dots. The observed lowering of the QD emission rate (by a factor of up to 1.7) gives a mechanism for greater control over the QD emission lifetime by a plasmonic nanoparticle than previously appreciated.

We show that the luminescence of QDs in the presence of a gold nanoparticle is enhanced. The steady-state emission is increased for the off-resonant dimer compared to the closely resonant dimer. Using an open quantum system-based analysis on extreme metal-QD dimer orientations, this is attributed to the dominance of axial enhancement of the electric field, and consequently the steady-state QD spectra. 

\section{Experimental}

\subsection{Materials}
Cadmium oxide (CdO) (99\%), octadecene  (ODE) (technical grade, 90\%), oleylamine (technical grade, 70\%), trioctylphosphine (90\%), oleic acid (technical grade, 90\%), octane thiol  (≥ 98.5\%), tetradecylphosphonic acid, selenium powder (Se) (99.00\%), sulfur (≥ 99.99\%), sodium borohydride (NaBH$_{4}$) (≥ 99\%), gold(III) chloride trihydrate ($\geq$ 99.9\%), tannic acid, tri-sodium citrate dihydrate (≥ 98.5\%), bis(\emph{p}-sulfonatophenyl)phenylphosphine (BSPP) (97\%), potassium carbonate, phosphonoacetic acid (98\%), agarose (type I, low EEO), boric acid, Tris base, chloroform (high-pressure liquid chromatography (HPLC) grade), EDTA and Ficoll 400 were purchased from Sigma Aldrich. CTAB (98\%) was purchased from Ajax Finechem, acetone (analytical reagent (AR) grade), and methanol (AR grade) were purchased from Merck. Thiolated ethylene glycol HS-C$_{11}$-(EG)$_{6}$-OCH$_{2}$-COOH was purchased from Prochimia Surface (Poland). Tetramethylammonium hydroxide (TMAOH) (25\% w/w in methanol) was purchased from Alfa Aesar.\\ 
PAGE-purified trithiolated DNA sequences were purchased from Fidelity Systems, Inc. (USA). The sequences of DNA used for DNA-based self-assembly are given below.

\textbf{A1:} \seqsplit{5'-trithiol-TTTTCTCACTAAGATCGATAGAGCGATTGTGATATTTCAAGCGGTACTCCAGCTCTAGGTAGCTCCCTTTCCAATCAGCTTATGTGAGCGCCTGCCCATG-3'}

\textbf{A1-comp:} \seqsplit{5'-trithiol-TTTCATGGGCAGGCGCTCACATAAGCTGATTGGAAAGGGAGCTACCTAGAGCTGGAGTACCGCTTGAAATATCACAATCGCTCTATCGATCTTAGTGAGA-3'}	

\textbf{B1:}	\seqsplit{5'-trithiol-TATACCTGACCTCGGGACTTGACTGATTGT-3'}	

\textbf{B1-comp:}	\seqsplit{5'-trithiol-ACAATCAGTCAAGTCCCGAGGTCAGGTATA-3'}

\textbf{L1:} \seqsplit{5'-GCTGACTCGCTACTCTTTTTTTTTTTTTTTTTTTTTTTTTTTTTTTTTTTTTTTTTTTTTTTTTTTTTTTTTTTTGCATGCAGATACAATCAGTCAAGTC-3'}	

\textbf{L1-comp:} \seqsplit{5'-GACTTGACTGATTGTATCTGCATGCTTTTTTTTTTTTTTTTTTTTTTTTTTTTTTTTTTTTTTTTTTTTTTTTTTTTTTTTTTTTGAGTAGCGAGTCAGC-3'}	

\textbf{L2:} \seqsplit{5'-AGCACACAAGAGCTGTTTTTTTTTTTTTTTTTTTTTTTTTTTTTTTTTTTTTTTTTTTTTTTTTTTTTTTTTTTTAGCAGATACATATACCTGACCTCGG-3'}	

\textbf{L2-comp:} \seqsplit{5'-CCGAGGTCAGGTATATGTATCTGCTTTTTTTTTTTTTTTTTTTTTTTTTTTTTTTTTTTTTTTTTTTTTTTTTTTTTTTTTTTTTCAGCTCTTGTGTGCT-3'}

All chemicals were used as received without further purification. Ultrapure water from a MilliQ system, with resistivity $>$18 M$\Omega$ was used for all aqueous solutions.

\section{Synthesis}
\textbf{Gold Nanoparticles:} Gold nanospheres were synthesised according to the method described by Piella \textit{et. al.}\cite{piella_2016_ChemMat}. Different aspect ratios of gold nanorods were synthesised using the method reported by Nikoobakht  and  El-Sayed.\cite{nikoobakht_2003_ChemMat}\\
\textbf{CdSe Core Nanocrystals:} CdSe cores were synthesized according to the reported method.\cite{carbone_2007_Nanolett}\\
\textbf{Shelling of CdSe Cores:} The CdSe core particles were shelled according to the published protocol of Boldt \textit{et. al.}\cite{boldt_2013_ChemMat}

\subsection{Ligand Exchange of Nanoparticles}
Quantum dots were transferred from the organic to aqueous phase by ligand exchange. For this, phosphonoacetic acid (PsAA) (0.1 M, 5 mL) solution was prepared in methanol and added to a dispersion of the QDs in chloroform (13 $\mu$M, 1.00 mL). The pH of the PsAA solution was maintained using TMAOH
(tetramethylaluminum hydroxide), (25\% solution in methanol). The QD dispersion in chloroform was added and then the dispersion was sonicated for 1-2 minutes. Water was added and the QDs migrated to the aqueous layer following sonication, leaving a colorless organic layer. The aqueous layer containing the QDs was collected and washed twice by precipitation via the addition of acetone ($\sim$ 5–10 mL).  The precipitate was centrifuged and the pellet was redispersed in water to remove excess PsAA.

Colloidal gold nanoparticles with citrate ligands with a diameter of 10 nm were mixed overnight with BSPP (200.0 $\mu$L, 90.0 mM). These were then purified by centrifugation at 13400 rpm for 50 min. The supernatant was removed and the pellet redispersed in BSPP buffer and washed three more times to achieve colloidally stable gold nanoparticles. After the final washing, the pellet was collected and re-dispersed in a known volume of BSPP to achieve a final concentration of nanoparticles (2 - 5 $\mu$M). 

\subsection{NP-DNA Functionalisation}

\textbf{QDs-Short (30 bp) DNA Conjugation:}
Thiolated short DNA (100.0 $\mu$M) (0.5, 1.0, 2.0, 3.0, 4.0, and 5.0 $\mu$L) were mixed with the equimolar concentration of non-functionalized lengthening DNA strands (100 bp) which was then left to incubate overnight. The DNA solution was then added to the highly concentrated quantum dot solution (25.0 $\mu$M, 5 $\mu$L) in the presence of BSPP (90.0 mM, 1.0 $\mu$L) and NaCl (1 M, 1 mL). The samples were incubated overnight before running on the gel for purification.  \\
\textbf{QDs-Long (100 bp) DNA Conjugation:}
Thiolated DNA (100.0 $\mu$M) (0.5, 1.0, 2.0, 3.0, 4.0, and 5.0 $\mu$L) were added to highly concentrated quantum dot solution (25.0 $\mu$M, 5 $\mu$L) in the presence of BSPP (90.0 mM, 1.0 $\mu$L) and NaCl (1 M, 1 mL). The samples were incubated overnight before running on the gel for purification. \\
\textbf{AuNP-Short (30 bp) DNA Conjugation:}
Short (30 bp) thiolated DNA (0.5, 1.0, 2.0, 3.0, 4.0, and 5.0 $\mu$L of 100.0 $\mu$M) were mixed with an equimolar concentration of non-functionalized lengthening DNA strands (100 bp) and the dispersion left to incubate overnight. The DNA solution was then added to the highly concentrated gold nanoparticle solution (2-5 $\mu$M, 5 $\mu$L) in the presence of BSPP (90.0 mM, 1.0 $\mu$L) and NaCl (1 M, 1 mL). The samples were incubated overnight before running on the gel for purification.\\
\textbf{AuNP-Long (100 bp) DNA Conjugation:}
The thiol-functionalised DNA (0.1, 1.0, 3.0 and 10.0 $\mu$L of 100.0 $\mu$M) were separately mixed with the highly concentrated dispersion of gold nanoparticles (AuNP) (2-5 $\mu$M, 5.0 $\mu$L) in the presence of salt (NaCl) (1.0 M, 1 mL) and  BSPP (90.0 mM, 1.0 $\mu$L). The samples were left overnight for incubation. 

\subsection{Purification of NP-DNA conjugates}

Prior to electrophoresis (15 minutes), mPEG ($\sim$100,000–500,000  molar excess with respect to the AuNP concentration, 1.0 $\mu$L) was added to the Au-DNA samples and HS-C$_{11}$-(EG)$_{6}$-OCH$_{2}$-COOH (340$X$ excess with respect to the QDs concentration, 1.0 $\mu$L) to the QD-DNA samples. Before loading the samples on the gel, Ficoll solution (20\%, 4.0 $\mu$L) was added as a loading buffer to increase the density of samples. The nanoparticle-DNA conjugate mixtures containing different concentrations of DNA were loaded on the agarose gel (2.7\% for AuNPs, 4.1\% for Q550, 3.8\% for Q570, 3.4\% for Q610, and 3.2\% for Q650). For the electrophoresis, the voltage was set to 80 V for 35 - 45 min. The same procedure was performed for all Au and QD particles, for both pairs of complementary DNA strands. An electro-elution procedure was used to extract AuNP-DNA conjugates from the gel and the extracted solution was concentrated via centrifugation before hybridization. For the extraction of QD-DNA conjugates, gel bands were cut and soaked in 0.5X TBE buffer overnight. The resulting dispersion was then concentrated by evaporating the solution under N$_{2}$ stream.

For COSA assemblies, the central AuNP was mixed with a higher DNA concentration (10-fold excess) to fully allow the DNA coverage around AuNP. The samples were left overnight for DNA conjugation and were then run on 1\% (w/v) of the agarose gel for purifying it from the unbound DNA strands.

For the preparation of gel, agarose powder was mixed with 0.5x TBE buffer to give the desired concentration of gel. The mixture was heated in the microwave oven (800 W) for 2 minutes to dissolve the gel. The gel was cooled to $\sim$60-70 $^{\circ}$C and poured into the electrophoresis tray. When the gel was fully solidified, the comb was removed. The gel was immersed in 0.5X TBE buffer. The sample was put into the wells created by the comb. Set the voltage to 80 V for 35 min and left the samples moved in the gel. 

\subsection{Self-Assembly and Purification}

After achieving purified samples containing the desired number of DNA strands per particle, AuNS and QDs with one DNA and one complementary DNA strand/particle respectively were mixed together in the presence of optimized salt concentration (1 M, 1.0 $\mu$L) to form the hybrid dimer assembly. Hybrid COSA assemblies were formed after mixing the central fully DNA functionalized AuNP cores with one complementary DNA containing QD satellite particles in the presence of salt (2 M, 1.0 $\mu$L).

The assemblies were run in the second gel electrophoresis step on a dilute gel (1\% - 2.5\%) to purify the dimer assembly from the unconjugated Au and QD particles and the satellite bearing core nanoparticles from the unbound satellites. The low mobility rate of dimers and core-satellites assemblies compared to single unconjugated nanoparticles was responsible for separate bands while running on the agarose gel. Therefore, the relative mobility rate forms the basis of purification of an assembly from the unconjugated nanoparticles. The band containing the purified structures of interest was cut from the gel, and the assembly samples were extracted separately from each band by the electro-elution procedure. The assembly solution was then concentrated upon centrifugation.

\subsection{Calculating Concentrations for Hybrid Assemblies}
We used inductively coupled plasma mass spectrometry (ICP-MS) to determine the concentrations of QDs in the hybrid assembly. An eight-point calibration curve was constructed using the known standard solutions of various concentrations. The correlation curves showed an r$_{2}$ value of 0.99. The concentration of the QDs was calculated using the isotopes of Cd(111) and Cd(112), Se (78), Se (82), Zn(66) Zn(68), and Au(197). 

\subsection{Sample Preparation}
Samples for electron microscopy were prepared by depositing 10 $\mu$L of the purified assembly on an ultra-thin TEM grid which was then soaked in a 1:1 mixture of water and ethanol for 10-15 minutes. The substrate was then rinsed with absolute ethanol and allowed to dry.
Quartz cuvettes of 3 mm path length were used for all photophysical measurements. The cuvette was pre-cleaned by soaking in a nitric acid bath overnight and then washing with copious amounts of water.

\section{Instrumentation}
Transmission electron microscopy (TEM) was performed on FEI Tecnai T20 Twin LaB6. Scanning electron microscopy was done on FEI Magellan 400 FEG SEM.\\
ICP-MS measurements were carried out using a PerkinElmer Nexion 350 ICP-MS instrument.\\
UV-Visible spectra were collected using an Agilent Cary 60 UV-Vis Spectrophotometer.\\ 
The luminescence spectra of the samples were recorded on a Cary Eclipse Fluorescence Spectrophotometer. For measuring the photoluminescence signal, the optical density of the samples was kept below 0.1 to avoid the inner filter effect.

\section{Photophysical Measurements}
The excitation and emission parameters used for the steady-state PL measurements were as follows: excitation wavelength = 425 nm, excitation slit = 5 nm, emission slit = 5 nm, scan rate = 600 nm/min and integration time = 0.1 s.\\
 
\subsection{Time Correlated Single Photon Counting}
The fluorescence lifetime decays of both the reference and the assembled quantum dots were measured via a home-built time-resolved single-photon counting (TCSPC) setup. Excitation was provided by a picosecond pulsed supercontinuum laser (Fianium, SC 400-4-pp) with the emission wavelength of 425 nm, selected using a 10 nm bandpass filter (laser power = 1 mW). The emission signal was collected at right angles to the excitation beam which was then directed towards the Multichannel plate detector. Photon emissions were recorded by a photon counting module (PicoHarp 300, Picoquant). The instrument response function (IRF) was measured by using a scattering solution of milk powder in water to be 0.6 ns. The TCSPC data was collected at the 5 MHz frequency and 32 ps time intervals.

\subsection{Data Analysis}
The time-resolved emission decays were fit with convolution with the IRF to a series of exponential functions until a good fit was obtained as judged by the reduced chi-squared (${\chi}^2$) value and the randomness of residuals. The fits were performed in custom-written routines in Igor Pro 6.3 software.

\begin{acknowledgement}

This work was undertaken within the ARC Centre of Excellence in Exciton Science, supported by the Australian Research Council (ARC) Grant CE170100026. The authors acknowledge the use of instruments and scientific and technical assistance at the Monash Centre for Electron Microscopy, a Node of Microscopy Australia. Computational resources were provided by the National Computational Infrastructure (NCI) at the Australian National University. HH gratefully acknowledges  Roslyn Forecast and Francesco Campaioli for insightful discussions.

\end{acknowledgement}

\begin{suppinfo}

Electron microscopy images of colloids, steady state spectra of all QDs and gold nanoparticles, wide-field images of assemblies, quantification of number of satellites in core-satellite assemblies, lifetime and steady-state data for assemblies with 100 bp DNA linkers and incorporating Q570 and Q650, along with tabulations of fit data are available online:

\end{suppinfo}

\bibliography{HybridAssemblies_Anum}

\end{document}


\setcounter{figure}{0}
        \renewcommand{\thefigure}{S\arabic{figure}}%
        \setcounter{table}{0}
        \renewcommand{\thetable}{S\arabic{table}}%

\begin{figure}[!ht]
    \includegraphics[width=\textwidth]{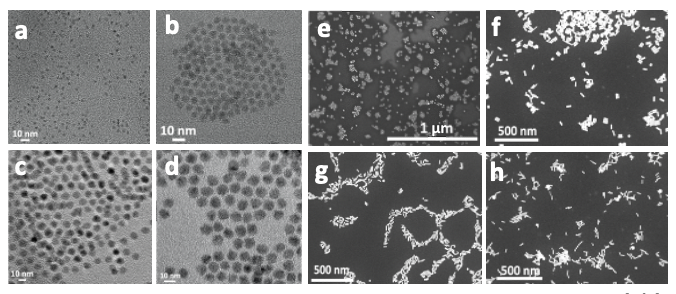}
\caption{Electron microscopy images of the nanocrystals used in the study. TEM images of (a) Q550 (b) Q570 (c) Q610 (d) Q650. SEM images of (e) AuNS$_{10}$ (f) AuNR$_{2.7}$ (g) AuNR$_{3.9}$ (h) AuNR$_{4.1}$.}
\label{EM-widefield_SI}
\end{figure}

\begin{figure}[!ht]
    \includegraphics[width=\textwidth]{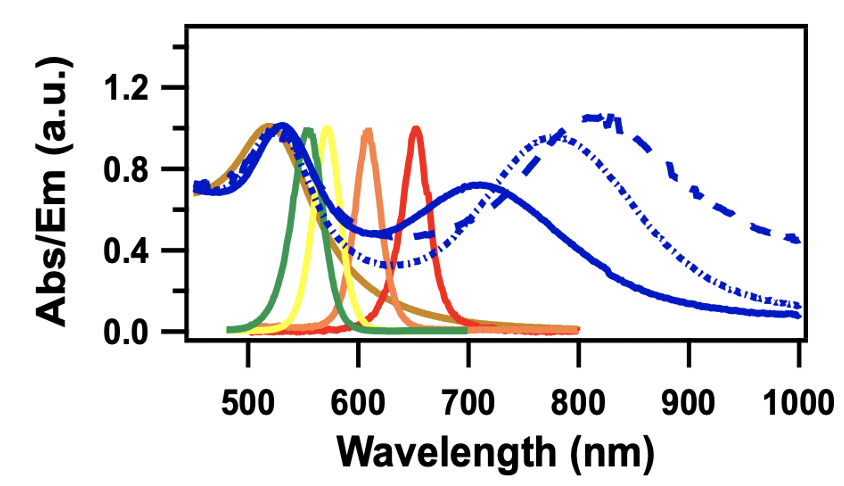}
\caption{Overlay of absorption spectra of AuNS$_{10}$, different aspect ratio of AuNR and the emission spectra of the quantum dots used in the optical study of hybrid assemblies; AuNS$_{10}$ (brown), AuNR$_{2.7}$ (line-blue), AuNR$_{3.9}$ (dotted-blue), AuNR$_{4.1}$ (dashed-blue), Q550 (green), Q570 (yellow), Q610 (orange) and Q650 (red).}
\label{AbsEm_SI}
\end{figure}

\begin{figure}[!ht]
    \includegraphics[width=\textwidth]{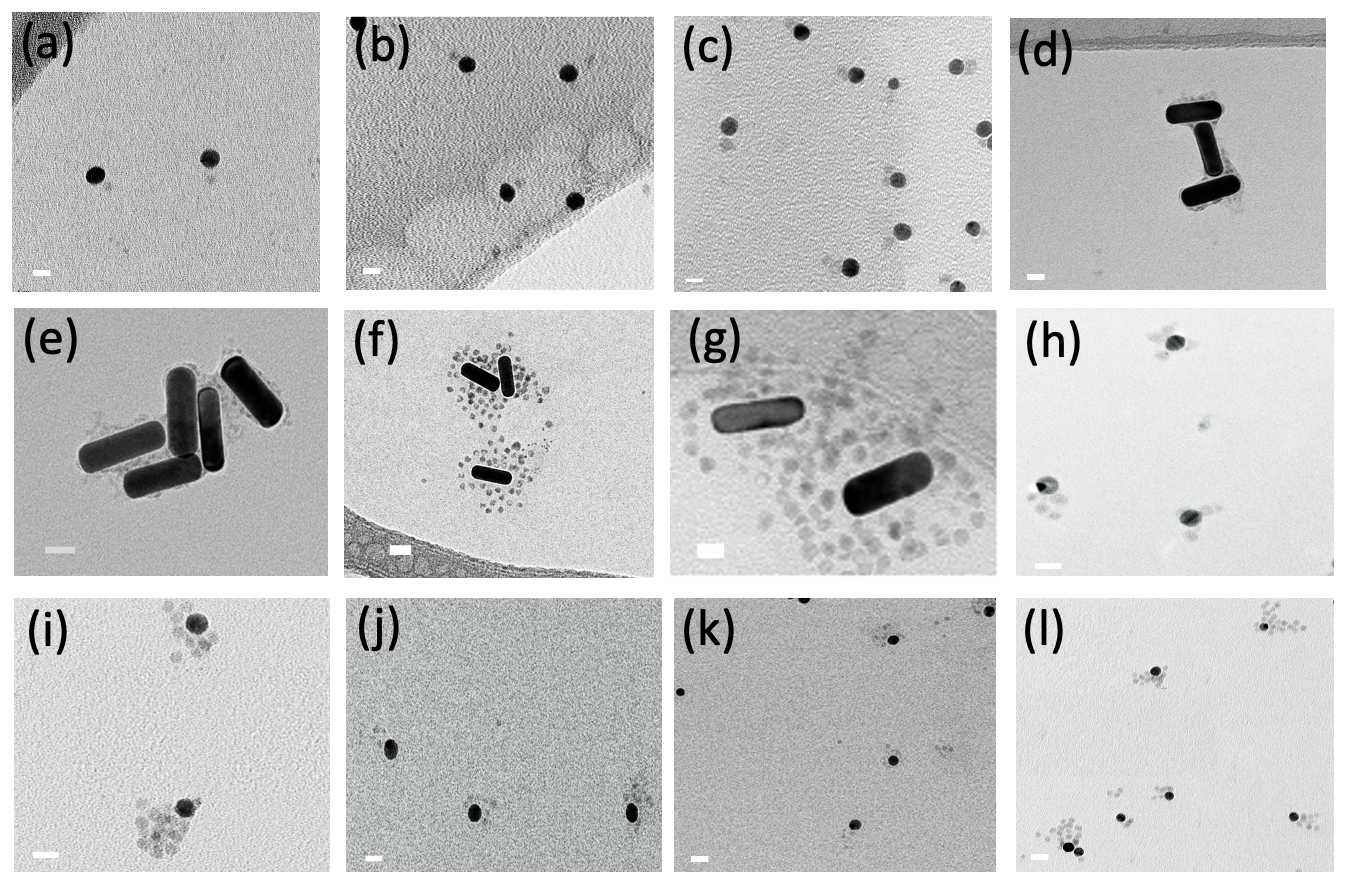}
\caption{Wide-field electron microscopy view of (a) Q550--AuNS$_{10}$ dimers with 100 bps DNA (b) Q550--AuNS$_{10}$ dimers with 100 bps DNA (c) Q610-AuNS$_{10}$ dimers with 30 bps DNA (d) Q550-AuNR$_{2.7}$-COSA with 30 bps DNA (e) Q570-$_{2.7}$-COSA with 30 bps DNA (f) Q610-AuNR$_{2.7}$ with 30 bps DNA (g) Q650-AuNR$_{2.7}$ with 100 bps DNA (h) Q610-AuNS$_{10}$ with 100 bps DNA (i) Q650-AuNS$_{10}$ with 30 bps DNA (j) Q550-AuNS$_{10}$ with 30 bps DNA (k) Q570-AuNS$_{10}$ with 30 bps DNA (l) Q610-AuNS$_{10}$ with 30 bps DNA. The scale bar for the TEM images is 10 nm.}
\label{EM-widefield2_SI}
\end{figure}

\begin{figure}[!ht]
    \includegraphics[width=\textwidth]{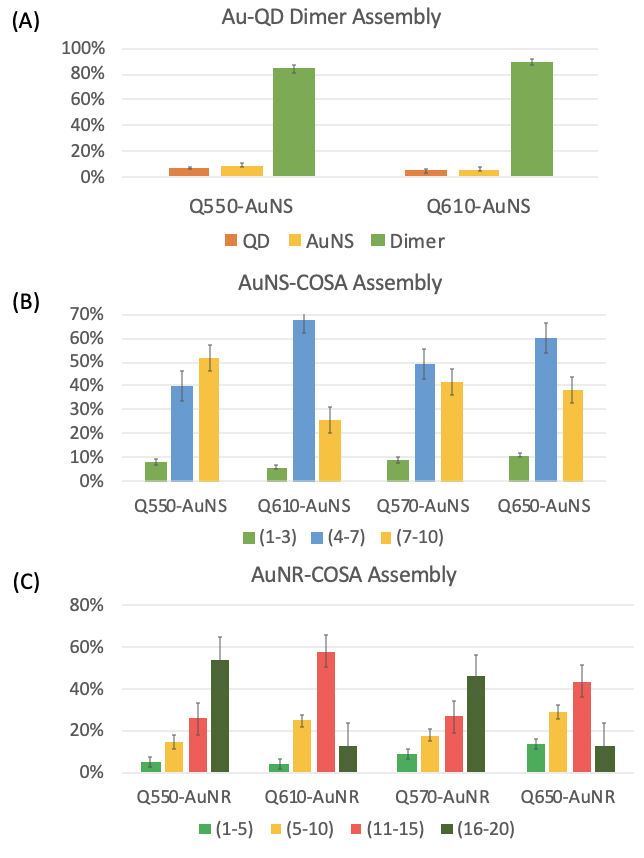}
\caption{Distribution of number of QDs around AuNP in the hybrid assemblies (A) Hybrid dimer assembly of Q550-AuNS$_{10}$ and Q610-AuNS$_{10}$ (B) Q550, Q570, Q610 and Q650-AuNS$_{10}$-COSA assembly (C) Q550, Q570, Q610 and Q650-AuNR$_{2.7}$-COSA assembly. The number of structures analysed from the TEM images are between 60-80 for each set of hybrid assembly.}
\label{QD_distribution}
\end{figure}

\begin{figure}[!ht]
    \includegraphics[width=\textwidth]{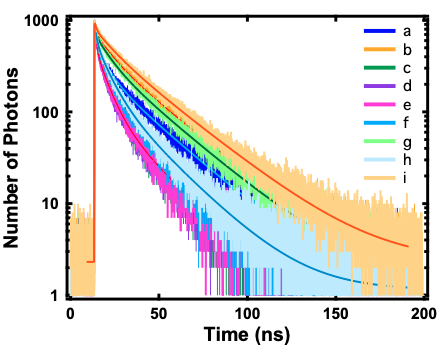}
\caption{Lifetime analysis of some of the controlled samples. a=Q610, b=Q610-
BSPP, c=Q610-BSPP-PEG, d=Q610-1DNA, e=Q610-full DNA, f=Q610-AuNS$_{10}$ (without DNA)-PEG, g=Q610-AuNS$_{10}$ (without DNA and PEG), h=Q610-AuNR$_{2.7}$-PEG (without
DNA)-PEG, i=Q610- AuNR$_{2.7}$ (without DNA and PEG) @ 425 nm, 32 ps resolution and 5 MHz frequency.}
\label{Control_exp_TCSPC}
\end{figure}

\begin{figure}[!ht]
    \includegraphics[width=\textwidth]{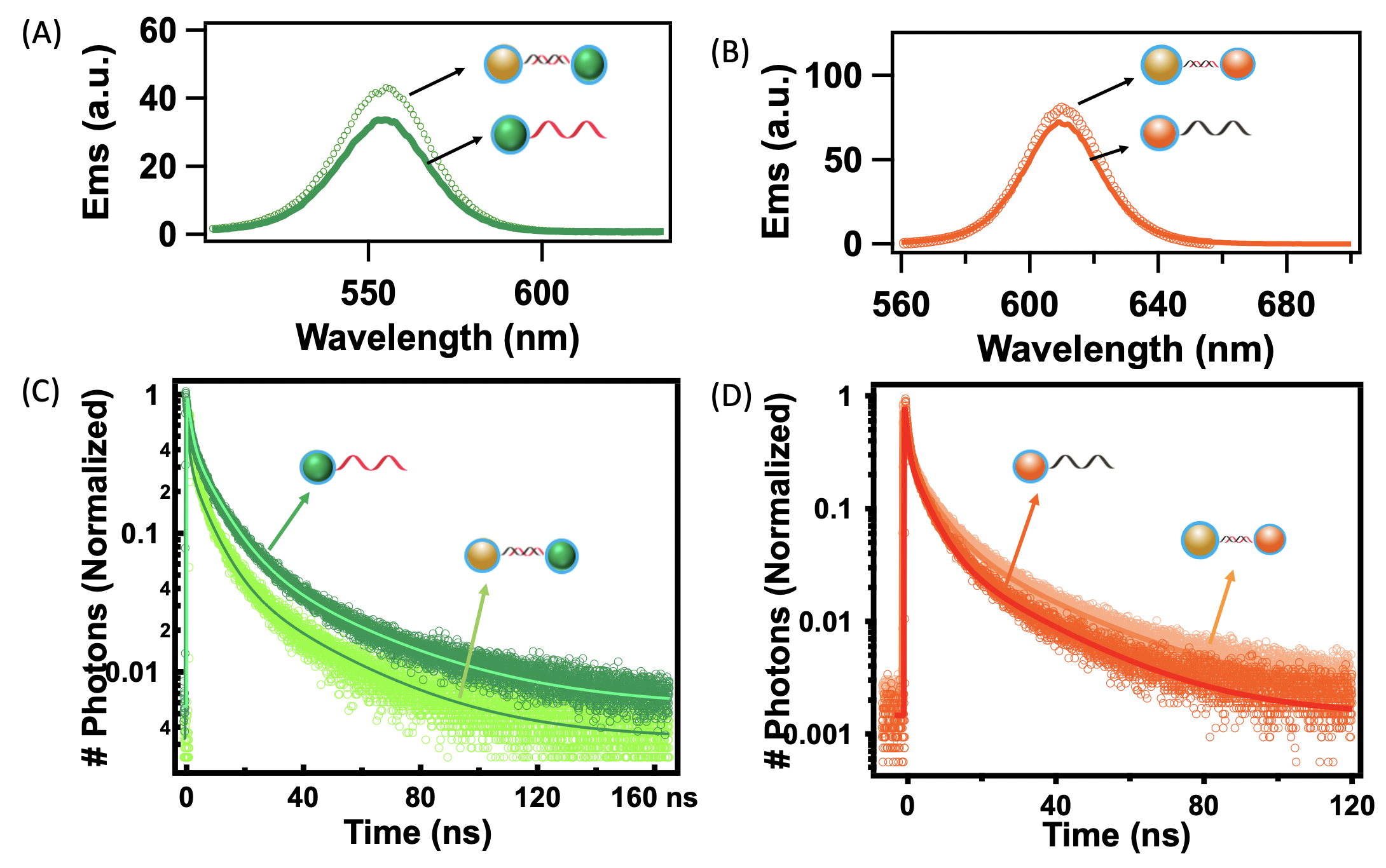}
\caption{Steady state fluorescence spectra of (A) Q550-Ref (dark green) and Q550-AuNS$_{10}$ dimers (light green) (B) Q610-Ref (dark orange) and Q610-AuNS$_{10}$ dimers (light orange) developed with 100 bps DNA with $\sim$ 34 nm interparicle separation. Lifetime decays of the hybrid dimer assemblies held together by 100 bps DNA @ 425 nm, 32 ps resolution and 5 MHz frequency (C) Q550-Ref (dark green) and Q550-AuNS$_{10}$ (light green) (D) Q610-Ref (dark orange) and Q610-AuNS$_{10}$ (light orange).}
\label{TCSPC_Dimers_Q550_Q610_LDNA}
\end{figure}

\begin{figure}[!ht]
    \includegraphics[width=\textwidth]{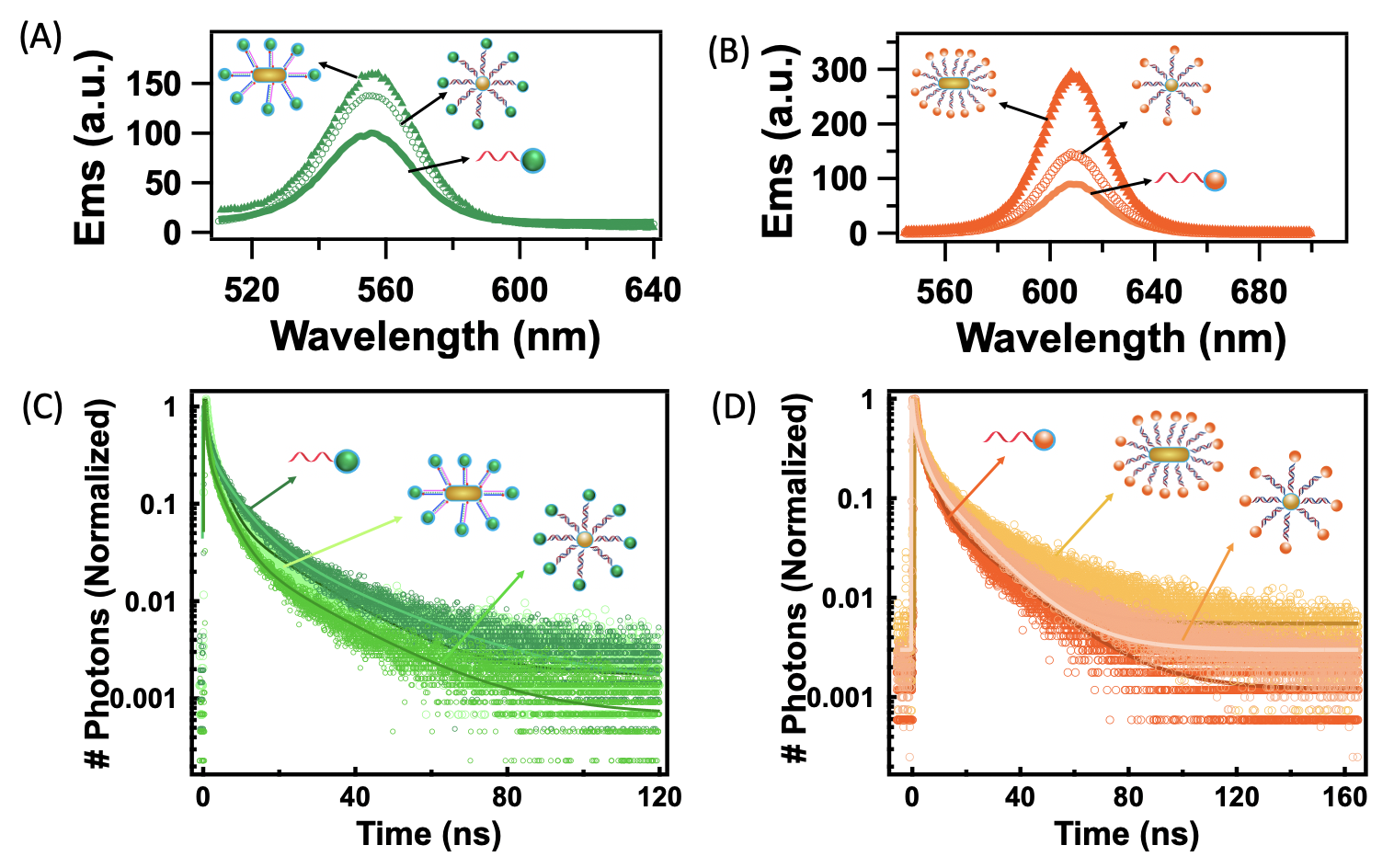}
\caption{Fluorescence emission spectra of the core-satellite hybrid constructs developed with 100 bps DNA ($\sim$ 34 nm interparicle separation) (A) Q550-Ref (plain green), Q550-AuNS$_{10}$ COSA (circle green), Q550-AuNR$_{2.7}$ COSA (traingle green) (B) Q610-Ref (plain orange), Q610-AuNS$_{10}$ COSA (circle orange), Q610-AuNR$_{2.7}$ COSA (triangle orange) assemblies. Normalized lifetime decays of core-satellites hybrid assemblies at an excitation wavelength of 425 nm, 32 ps resolution and 5 MHz frequency. (C) Q550-Ref (dark green), Q550-AuNS$_{10}$ COSA (medium green) and Q550-AuNR$_{2.7}$ COSA (light green) (D) Q610-Ref (dark orange), Q610-AuNS$_{10}$ COSA (medium orange), Q610-AuNR$_{2.7}$ COSA (light orange) assemblies.}
\label{TCSPC_COSA_Q550_Q610_LDNA}
\end{figure}

\begin{figure}[!ht]
    \includegraphics[width=\textwidth]{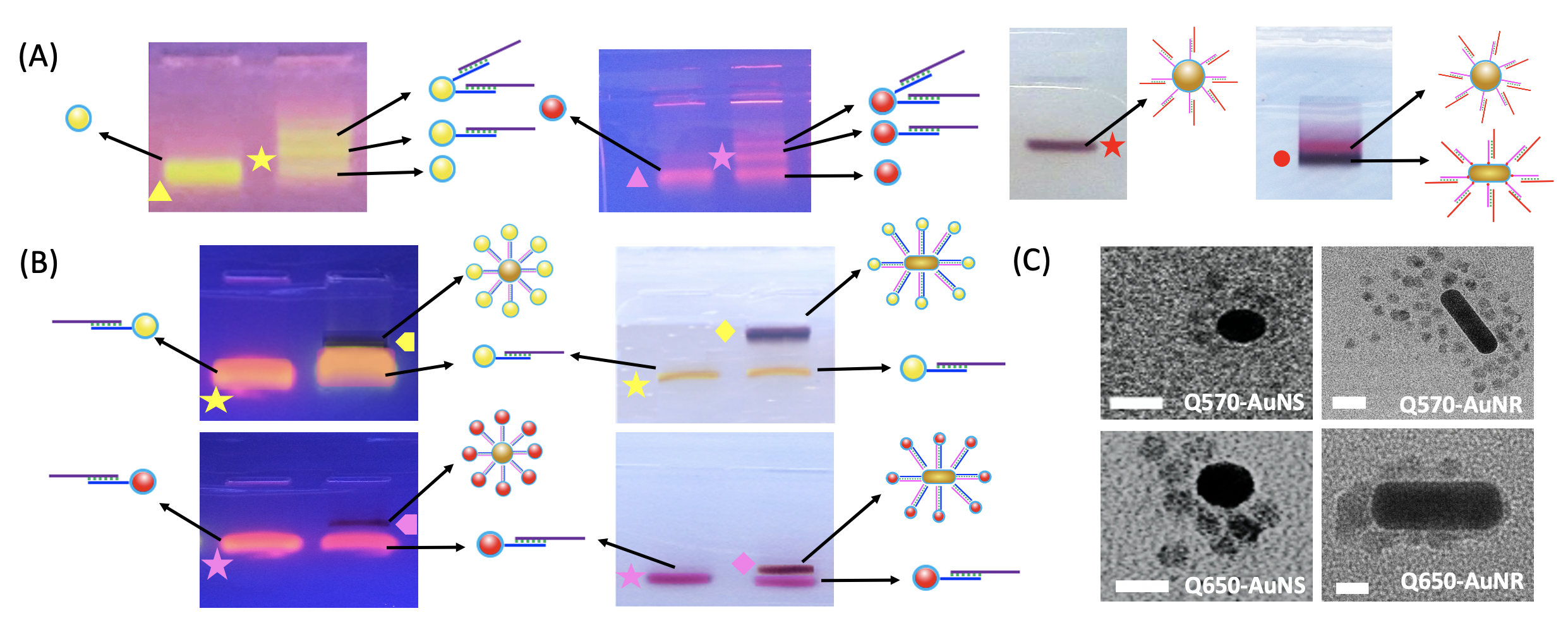}
\caption{
(A) Optimization of 30 bps DNA concentration for achieving 1-DNA/particle for Q570 (yellow star) and Q650 (pink star) and fully functionalized AuNS$_{10}$ (red star) and AuNR$_{2.7}$ (red circle) particles. (B) Purification of hybrid Q570-AuNS$_{10}$-COSA (yellow pointer), Q570-AuNR$_{2.7}$-COSA (yellow diamond), Q650-AuNS$_{10}$-COSA (pink pointer), Q650-AuNR$_{2.7}$-COSA (pink diamond) assemblies from the unconjugated Q570 and Q650 (C) TEM characterization of the synthesized Q570 and Q650 COSA assemblies. The average number of satellites are $\sim$4-7 for Q570-AuNS$_{10}$ and Q650-AuNS$_{10}$ COSA assemblies while $\sim$16-20 and $\sim$11-15 for Q550-AuNR$_{2.7}$ COSA and Q650-AuNR$_{2.7}$ COSA assemblies, respectively.}
\label{Gels_Q570_Q650}
\end{figure}

\begin{figure}[!ht]
    \includegraphics[width=\textwidth]{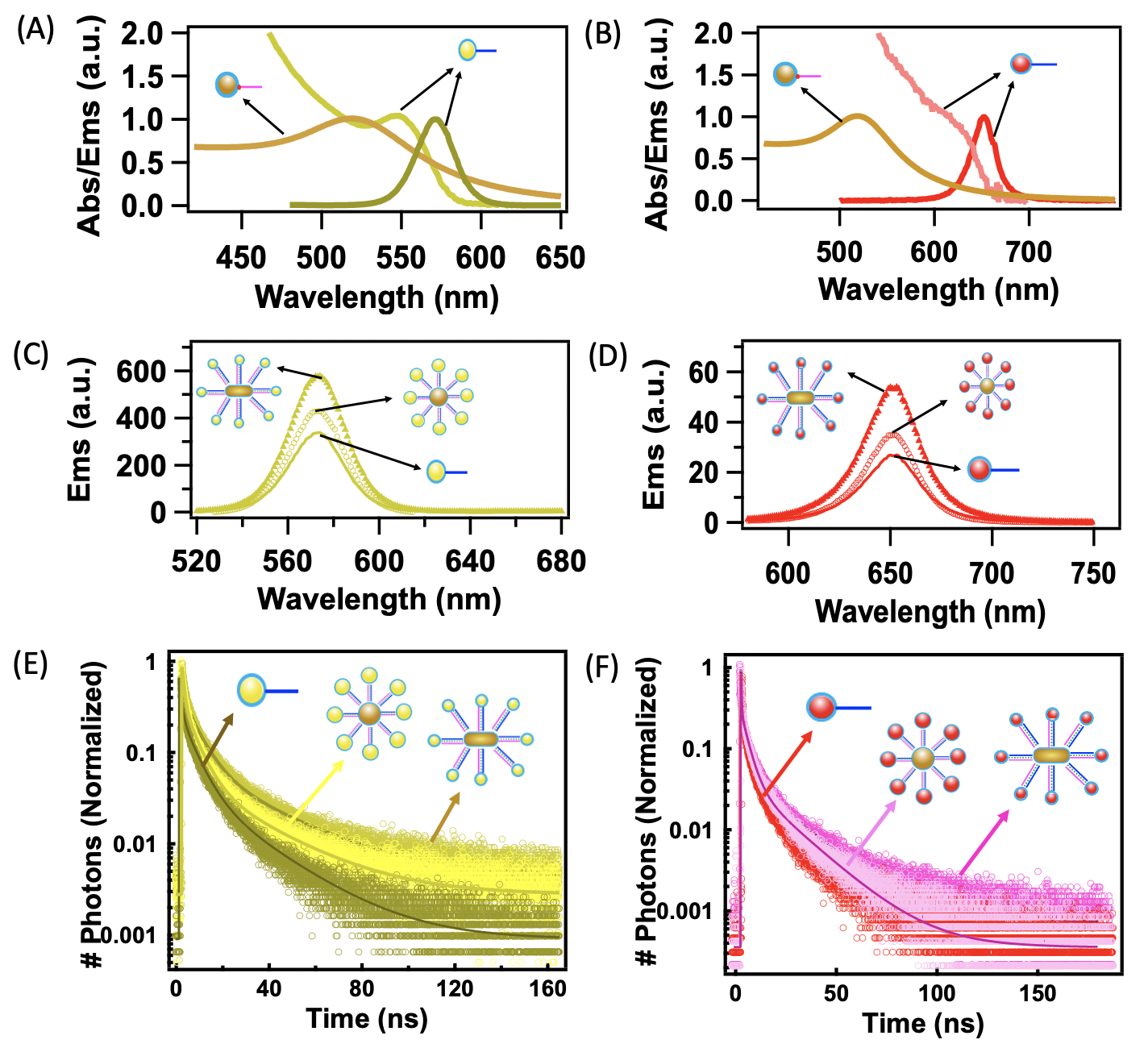}
\caption{(A,B) Absorption spectrum of AuNS$_{10}$ (brown) overlaped with the absorption and emission spectra of Q570 (light and dark dirty green) and Q650 (light and dark red). Steady-state PL spectra of (C) Q570-Ref (plain dirty green), Q570-AuNS$_{10}$-COSA (circle dirty green), and Q570-AuNR$_{2.7}$-COSA (triangle dirty green) (D) Q650-Ref (plain red), Q650-AuNS$_{10}$-COSA (circle red), and Q650-AuNR$_{2.7}$-COSA (triangle red) hybrid assemblies. Lifetime decay traces of (E) Q550-Ref (dark dirty green), Q570-AuNS$_{10}$-COSA (yellow), Q570-AuNR$_{2.7}$-COSA (light dirty green) (F) Q650-Ref (red), Q650-AuNS$_{10}$-COSA (light pink), Q650-AuNR$_{2.7}$-COSA (dark pink) hybrid assemblies @ 425 nm, 32 ps resolution and 5 MHz frequency.}
\label{Photophysics_COSA_Q570_Q650}
\end{figure}

\clearpage

\begin{table}
  \caption{Fitting parameters for the lifetimes of the control samples extracted at various stages of gel electrophoresis.}
  \label{Fit parameters for tunable LSPR}
  \begin{tabular}{ccccc}
    \hline
    Samples & $\tau_{1}(A_{1}\%)$ & $\tau_{2}(A_{2}\%)$  & $\tau_{3}(A_{3}\%)$ & $\langle\tau\rangle$ \\
    & ($\SI{}{\nano\second}$) & ($\SI{}{\nano\second}$) & ($\SI{}{\nano\second}$) & ($\SI{}{\nano\second}$) \\
    \hline
    Q610-Organic ligands & 0.6 (46.3) & 6.2 (26.5) & 25.6 (27.1) & 21.2  \\
    Q610-BSPP  & 0.9 (28.1) & 8.1 (27.9) & 27.4 (43.8) & 23.9 \\
    Q610-BSPP/PEG  & 0.8 (51.9) & 6.0 (36.4) & 21.7 (11.6) & 13.3 \\
    Q610-1DNA/PEG  & 0.9 (52.5) & 6.3 (36.7) & 22.3 (10.6) & 13.1 \\
    Q610-full DNA/PEG  & 0.8 (52.0) & 6.1 (36.7) & 21.8 (11.1) & 13.1\\
    Q610-AuNS$_{10}$ (No DNA)/PEG  & 0.9 (40.8) & 6.8 (41.5) & 21.8 (17.6) & 14.6 \\
    Q610-AuNR$_{2.7}$ (No DNA and PEG)  & 0.9 (19.3) & 9.9 (25.7) & 28.4 (54.9) & 25.6 \\
    
    \hline
\end{tabular}
\end{table}

\begin{table}
  \caption{Fitting parameters for the multiexponential decays of the QDs in hybrid assemblies fabricated using 30 bps DNA.}
  \label{Fit parameters 30bp}
  \begin{tabular}{ccccc}
    \hline
    Samples & Assembly & $\tau_{1}(A_{1}\%)$ & $\tau_{2}(A_{2}\%)$ & $\tau_{3}(A_{3}\%)$\\
    & Type & ($\SI{}{\nano\second}$) & ($\SI{}{\nano\second}$) & ($\SI{}{\nano\second}$) \\
    \hline
    Q550-Ref   & - & 0.1 (51.6) & 2.6 (30.3) & 14.4 (17.9)  \\
    Q550-AuNS$_{10}$ & Dimer & 0.1 (77.3) & 1.1 (17.3) & 10.6 (5.2)  \\
    Q550-Ref  & - & 1.0 (61.6) & 5.8 (32.4) & 23.0 (5.9)  \\
    Q550-AuNS$_{10}$  & COSA & 0.5 (71.7) & 3.7 (25.0) & 18.9 (3.1)  \\
    Q550-AuNR$_{2.7}$  & COSA & 0.3 (79.0) & 2.8 (18.9) & 17.5 (2.0)  \\
    Q570-Ref  & - & 0.6 (61.1) & 4.9 (31.1) & 22.9 (7.7)  \\
    Q570-AuNR$_{2.7}$  & COSA & 1.1 (54.1) & 7.1 (36.7) & 28.4 (9.1)  \\
    Q570-AuNS$_{10}$  & COSA & 0.9 (54.9) & 6.3 (36.5) & 26.0 (8.5)  \\
    Q610-Ref  & - & 0.9 (60.3) & 6.2 (32.9) & 23.7 (6.7)  \\
    Q610-AuNS$_{10}$  & Dimer & 0.9 (57.4) & 6.7 (35.4) & 30.9 (7.0)  \\
    Q610-Ref  & - & 1.0 (53.5) & 6.2 (37.2) & 21.4 (9.2)  \\
    Q610-AuNR$_{2.7}$  & COSA & 1.2 (33.9) & 10.1 (38.3) & 26.8 (27.7)  \\
    Q610-AuNS$_{10}$  & COSA & 0.9 (41.9) & 8.7 (33.6) & 26.1 (24.3)  \\
    Q650-Ref  & - & 0.2 (76.3) & 3.2 (20.4) & 18.3 (3.1)  \\
    Q650-AuNR$_{2.7}$  & COSA & 0.2 (75.8) & 3.2 (20.6) & 18.4 (3.5)  \\
    Q650-AuNS$_{10}$  & COSA & 0.3 (72.8) & 3.7 (23.4) & 21.1 (3.7)  \\
    \hline
\end{tabular}
\end{table}

\begin{table}
  \caption{Fitting parameters for the QDs in hybrid dimer assemblies fabricated using 100 bps DNA.}
  \label{Fit parameters of dimers 100 bps}
  \begin{tabular}{cccccc}
    \hline
    Samples & Assembly & $\tau_{1}(A_{1}\%)$ & $\tau_{2}(A_{2}\%)$ & $\tau_{3}(A_{3}\%)$ & $\langle\tau\rangle$ \\
    & Type & ($\SI{}{\nano\second}$) & ($\SI{}{\nano\second}$) & ($\SI{}{\nano\second}$) & ($\SI{}{\nano\second}$) \\
    \hline
    Q550-Ref   & - & 1.9 (60.3) & 7.1 (32.9) & 18.0 (6.7) & 8.6  \\
    Q550-AuNS$_{10}$ & Dimer & 0.9 (57.4) & 6.1 (35.4) & 12.8 (7.0) & 7.0 \\
    Q610-Ref  & - & 0.9 (78.1) & 12.0 (14.5) & 19.8 (7.2) & 11.8  \\
    Q610-AuNS$_{10}$  & COSA & 1.0 (83.0) & 13.9 (12.0) & 19.9 (4.8) & 14.3  \\
    \hline
\end{tabular}
\end{table}

\begin{table}
  \caption{Fitting parameters for the QDs in hybrid COSA assemblies developed by 100 bps DNA.}
  \label{Fit parameters COSA 100bp}
  \begin{tabular}{cccccc}
    \hline
    Samples & Assembly & $\tau_{1}(A_{1}\%)$ & $\tau_{2}(A_{2}\%)$  & $\tau_{3}(A_{3}\%)$ & $\langle\tau\rangle$\\
    & Type & ($\SI{}{\nano\second}$) & ($\SI{}{\nano\second}$) & ($\SI{}{\nano\second}$) & ($\SI{}{\nano\second}$) \\
    \hline
    Q550-Ref   & - & 1.5 (61.5) & 5.0 (32.4) & 20.0 (5.9) & 8.8  \\
    Q550-AuNS$_{10}$ & COSA & 0.5 (71.7) & 3.7 (25.0) & 18.9 (3.1) & 7.8 \\
    Q550-AuNR$_{2.7}$  & COSA & 0.3 (79.0) & 2.8 (18.9) & 17.5 (2.0) & 6.7 \\
    Q610-Ref  & - & 0.4 (83.1) & 3.4 (12.4) & 20.8 (4.4) & 12.2  \\
    Q610-AuNS$_{10}$  & COSA & 0.7 (65.9) & 5.6 (28.8) & 26.5 (5.1) & 13.0 \\
    Q610-AuNR$_{2.7}$  & COSA & 0.7 (57.1) & 6.4 (32.2) & 24.8 (10.5) & 15.3 \\
    \hline
\end{tabular}
\end{table}

\begin{table}
  \caption{Fitting parameters for the QDs lifetimes in hybrid assemblies with tunable LSPR.}
  \label{Fit parameters tunable LSPR}
  \begin{tabular}{cccccc}
    \hline
    Samples & Assembly & $\tau_{1}(A_{1}\%)$ & $\tau_{2}(A_{2}\%)$ & $\tau_{3}(A_{3}\%)$ & $\langle\tau\rangle$ \\
    & Type & ($\SI{}{\nano\second}$) & ($\SI{}{\nano\second}$) & ($\SI{}{\nano\second}$) & ($\SI{}{\nano\second}$) \\
    \hline
    Q610-Ref   & - & 1.0 (53.5) & 6.2 (37.2) & 21.4 (9.2) & 11.8 \\
    Q610-AuNS$_{10}$  & COSA & 0.9 (41.9) & 8.7 (33.6) & 26.1 (24.3) & 19.8 \\
    Q610-AuNR$_{2.7}$  & COSA & 1.2 (33.9) & 10.1 (38.3) & 26.8 (27.7) & 20.3 \\
    Q610-AuNR$_{3.9}$  & COSA & 1.0 (43.5) & 8.2 (32.6) & 25.0 (23.5) & 18.8 \\
    Q610-AuNR$_{4.1}$ & COSA & 0.8 (48.6) & 6.8 (34.0) & 22.6 (17.2) & 15.8 \\
    \hline
\end{tabular}
\end{table}

\clearpage


\section*{Supplementary information on theory}
\subsubsection*{Mie theory based modelling of modified spontaneous emission}

\begin{figure}[t!]  
	\includegraphics[width=0.5\columnwidth]{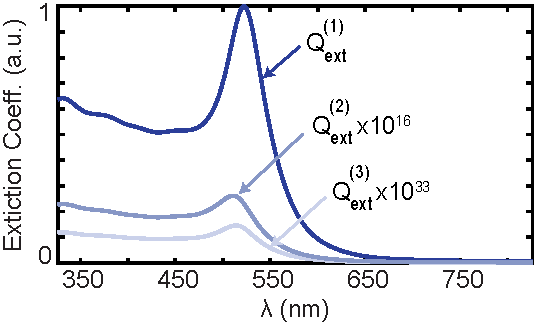}
	\centering
	\caption{Estimated extinction efficiencies for the dipolar (n=1), quadrupolar (n=2) and octupolar (n=3) modes of a gold nanoparticle of radius $\sim \SI{5}{\nano\meter}$ in water. Quadrupolar and octupolar spectra are magnified $10^{16}$ and $10^{33}$ times, respectively. \label{Fig:extinction_coefficient}}
\end{figure}

It has been shown that for very small centre separations ($z_0$) between a spherical metal nanoparticle and a dipolar emitter where $k z_0 \ll 1$ ($k$ is the wavenumber), the normalized emitter decay rate can be estimated using Mie theory, while considering all plasmon modes of the metal nanoparticle as follows \cite{colas2012mie}:

\begin{subequations}
\begin{align}
	\frac{\gamma^\perp}{n_\text{b}\gamma_0} = \frac{\tau_\text{ref}}{\tau^\perp} &\approx \frac{3}{2}\frac{1}{(k_\text{b} z_0)^3}\sum_{n=1}^\infty\frac{(n+1)^2}{z_0^{(2n+1)}}\Im\left[\alpha_n^\text{eff}\right] \label{Eq:gamma_perp}\\
	\frac{\gamma^\parallel}{n_\text{b}\gamma_0}  = \frac{\tau_\text{ref}}{\tau^\parallel} &\approx \frac{3}{4}\frac{1}{(k_\text{b} z_0)^3}\sum_{n=1}^\infty\frac{n(n+1)}{z_0^{(2n+1)}}\Im\left[\alpha^\text{eff}_n\right].\label{Eq:gamma_parallel}
\end{align}
\end{subequations}
In the above equations, $\gamma^\perp = 1/\tau^\perp $, $\gamma^\parallel = 1/\tau^\parallel$ and $\gamma_0$ denote the spontaneous emission rates of dipole emitter orientations perpendicular and parallel to the MNP surface, and  free-space spontaneous decay rate, respectively. The lifetime (decay time) of an isolated emitter in a medium of refractive index $n_\text{b}$ is denoted by $\tau_\text{ref}$. The decay rate of a randomly oriented emitter $\gamma$ can be estimated as $\gamma = (\gamma^\perp + 2\gamma^\parallel)\big/3$. The effective polarizability of the metal nanoparticle can be obtained as,
\begin{equation}
	\alpha_n^\text{eff} = \left[1-i\frac{(n+1)k_\text{b}^{(2n+1)}}{n(2n-1)!!(2n+1)!!}\alpha_n\right]^{-1}\alpha_n,
\end{equation}
where $\alpha_n$ is the polarizability attributable to the $n^\text{th}$  multipole tensor moment $\bm{p}^{(n)}$ of the metal nanoparticle, for an incident electric field $\bm{E}_0$ given by,
\begin{subequations}
	\begin{align}
		\bm{p}^{(n)}&=\frac{4\pi\epsilon_0\epsilon_\text{b}}{(2n-1)!!}\alpha_n\nabla^{n-1}\bm{E}_0\\
		\alpha_n &=\frac{n(\epsilon_\text{m} - \epsilon_\text{b})}{n\epsilon_\text{m} + (n+1)\epsilon_\text{b}}r_\text{m}^{(2n+1)}. \label{Eq:alpha_n}
	\end{align}
\end{subequations}
The relative permittivity of the metal nanoparticle and the absolute permittivity of free-space are denoted by $\epsilon_\text{m}$ and $\epsilon_0$, respectively. The double factorials in the above equation are computed using the pattern $(2n+1)!! = 1\times3\times5\times...\times(2n+1)$ and $\nabla$ denotes the vector differential operator. To decide the number of modes to be considered in the infinite summations in equations (\ref{Eq:gamma_perp}) and (\ref{Eq:gamma_parallel}), we evaluated the extinction efficiency of the metal nanoparticle for a few different modes, as shown in Figure\ \ref{Fig:extinction_coefficient}.

\begin{figure}[t!]  
	\includegraphics[width=0.5\columnwidth]{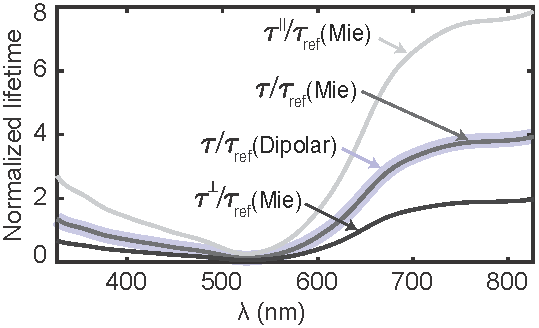}
	\centering
	\caption{Estimated lifetimes for generic emitters near MNPs calculated using equations (\ref{Eq:gamma_perp}) and  (\ref{Eq:gamma_parallel}).}  \label{Fig:Mie_decay}
\end{figure}

When obtaining Figure\ \ref{Fig:extinction_coefficient}, extinction efficiency of the $n^\text{th}$ plasmon mode of the metal nanoparticle at a given angular frequency $\omega$ was modelled as $Q_\text{ext}^{(n)}(\omega) = 4 k \Im\left[\alpha_n^\text{eff}\right]\big/r_\text{m}^2$ \cite{colas2012mie}. The permittivity of the gold nanoparticle was obtained using the tabulations by Johnson and Christy \cite{johnson1972optical}. We can readily observe that the extinction efficiency of the dipolar ($n=1$) mode exceeds the quadrupolar (n=2) and octupolar ($n=3$) modes by factors in the order of $10^{16}$ and $10^{33}$. Therefore in our case, multipolar effects can be safely neglected, taking only the dipole plasmon mode into account. We could then obtain the lifetime spectra of a generic dipolar emitter kept next to an MNP modelled using the dipole term ($n=1$) in the Mie expansion depicted in Figure\ \ref{Fig:Mie_decay}, using equations (\ref{Eq:gamma_perp}) and (\ref{Eq:gamma_parallel}). 

\subsubsection*{Nonlocally modelling the metal nanoparticle}

Let us now investigate the impact of replacing $\alpha_n^\text{eff}$ in the two equations (\ref{Eq:gamma_perp}) and (\ref{Eq:gamma_parallel}) by the local dipolar polarizability $\alpha_1$ obtained by setting $n=1$ in equation (\ref{Eq:alpha_n}). Normalized lifetime spectra obtained in this manner for a randomly oriented emitter placed at a centre separation $z_0 \sim \SI{17}{\nano\meter}$ from the MNP is shown by the purple curve in Figure\ \ref{Fig:Mie_decay}. This curve coincides with the earlier normalized lifetime curve for a random emitter obtained using the effective polarizability, displaying the sufficiency of using the dipolar polarizability $\alpha_1$ in place of $\alpha_n^\text{eff}$ in the normalized decay rate equations (\ref{Eq:gamma_perp}) and (\ref{Eq:gamma_parallel}). 

Mie expansion overlooks the nonlocal effects \cite{raza2015nonlocal, mortensen2014generalized} that become quite prominent in small MNPs with $\sim\SI{10}{\nano\meter}$ diameter \cite{raza2013blueshift}, where the ratio of the number of surface atoms to those that make up the bulk of the particle is significant \cite{artuso2012optical}. Local modelling of the MNP optical response has been challenged on a number of accounts, such as claims that the surface plasmon resonance energy in the quasistatic limit is independent of the MNP size, which conflict with observations in plasmonic experiments \cite{raza2015nonlocal, tiggesbaumker1993blue, wubs2017nonlocal, raza2013blueshift, raza2013blueshift2}. Thus, we use the recently introduced generalized nonlocal optical response (GNOR) theory \cite{raza2015nonlocal} that has successfully explained plasmonic experiments to model the small MNP in our MNP-QE nanohybrid model.

The GNOR model is a recent generalization and an extension of the nonlocal hydrodynamic (HDM) model \cite{raza2015nonlocal}, which goes beyond HDM by taking both convection current and electron diffusion phenomena in the MNPs into account \cite{wubs2017nonlocal, raza2015nonlocal}. We can convert the local dipolar polarizability $\alpha_1$ into its nonlocal form  $\alpha_1^\text{NL}$ by incorporating the GNOR correction factor $\delta_\text{NL}$ as follows,

\begin{equation}\label{Eq:NL_alpha_l}
\alpha_1^\text{NL} = \frac{\epsilon_\text{m}(\omega) - \epsilon_\text{b}(1+\delta_{\text{NL}})}{\epsilon_\text{m}(\omega)+2\epsilon_\text{b}(1+\delta_{\text{NL}})}r_\text{m}^3.
\end{equation}
where the GNOR correction factor $\delta_{\text{NL}}$ is given by,
\begin{equation}\label{Eq:delta_NL}
\delta_{\text{NL}} = \frac{\epsilon_\text{m}(\omega) - \epsilon_\infty(\omega)}{\epsilon_\infty(\omega)}\frac{j_1(K_\text{L} r_\text{m})}{K_\text{L} r_\text{m} j_1' (K_\text{L} r_\text{m})}.
\end{equation}
In equation (\ref{Eq:delta_NL}), $j_1$ denotes the spherical Bessel function of the first kind of angular momentum order 1, $j_1'$ denotes its first order differential with respect to the argument. The longitudinal wave vector $K_\text{L}$ abides by the relationship $K_\text{L}^2 = \epsilon_\text{m}(\omega)/\xi^2\left(\omega\right)$. This model obtains the bound electron response of the the MNP using experimentally measured bulk dielectric data ($\epsilon_{\text{expt}}$) as, $\epsilon_\infty = \epsilon_{\text{expt}}(\omega) + \omega_\text{p}^2\big/\left[\omega(\omega+i\Gamma_\text{abs})\right]$ whereas the nonlocal parameter of the GNOR model is characterized by, 
\begin{equation}
\xi^2(\omega) = \frac{\epsilon_\infty(\omega) \left[\kappa^2 + \mathrm{D}\left(\Gamma_\text{abs} - i\omega\right)\right]}{\omega\left(\omega+i\Gamma_\text{abs}\right)},
\end{equation}
where D is the electron diffusion constant and $\kappa^2 = \left(3\big/5\right)v_F^2$ for $\omega \gg \Gamma_\text{abs}$ (in the high frequency limit) and $v_F$ is the Fermi velocity of the MNP. It is evident that we can retrieve the locally modelled optical response $\alpha_1$ by setting $\delta_\text{\tiny{NL}}\to 0$ in (\ref{Eq:NL_alpha_l}). We will be using the the nonlocally corrected dipolar optical response $\alpha_1^\text{NL}$ to theoretically characterize the small MNP in our nanohybrid, from this point onwards.

We insert (\ref{Eq:NL_alpha_l}) in place of $\alpha_n^\text{eff}$ in (\ref{Eq:gamma_perp}) and (\ref{Eq:gamma_parallel}) and set $n=1$ to obtain the nonlocally corrected versions of the normalized emitter decay rate equations for dipolar emitters placed near small MNPs presented in the main text.
  
\subsubsection*{Modelling the QD exciton as a two-level open quantum system}\label{Sec:TLA}

We can obtain the Hamiltonian of a two-level atomic (TLA) system under the influence of the MNP and the externally applied electric field as \cite{hapuarachchi2020influence, hapuarachchi2018exciton, artuso2012optical},
\begin{equation}\label{Eq:QD_Hamiltonian}
\hat{\mathcal{H}} = \hbar\omega_0\hat{\sigma}^+\hat{\sigma}^- - \mu E_\text{qd}(\hat{\sigma}^+ +\hat{\sigma}^-),
\end{equation}
where $\hbar\omega_0$ is the excitonic energy (matched to the QD emission peak), and $E_\text{qd}$ is the total electric field incident on the QD exciton. The operators and states $\hat{\sigma}^- = |g\rangle \langle e|$, $\hat{\sigma}^+ = |e\rangle \langle g|$, $|g\rangle = (1,0)^\text{T}$ and $|e\rangle = (0,1)^\text{T}$ denote the exciton annihilation and creation operators, and the atomic ground and excited states, respectively. We can obtain $E_\text{qd}$ as \cite{hapuarachchi2020influence, hapuarachchi2018exciton, artuso2012optical},

\begin{subequations}
	\begin{align}
	&E_\text{qd} = \tilde{E}^+_\text{qd}e^{-i\omega t} + c.c.,\label{Eq:E_qd}\\
	&\tilde{E}^+_\text{qd} = \frac{\hbar}{\mu}\Omega^r = \tilde{E}^+_1 + \tilde{E}^+_2 + \tilde{E}^+_3,\\	\label{Eq:E_qd_tilde_plus}
	&\tilde{E}^+_1 = \frac{E_0}{\epsilon_\text{effS}}, \text{\tab} \tilde{E}_2^+= \frac{s_\alpha E_0 \alpha_1^\text{NL}}{ \epsilon_\text{effS} z_0^3}, \text{\tab} \tilde{E}_3^+ = \frac{s_\alpha^2 \mu \alpha_1^\text{NL}}{\left(4\pi\epsilon_0\epsilon_\text{b}\right)z_0^6 \epsilon_\text{effS}^2}\tilde{\rho}_\text{eg}
	\end{align}
\end{subequations}

In the above equations, $\tilde{E}^+_\text{x}$ denotes the slowly-time-varying positive frequency amplitude of the respective $E_\text{x}$, $\omega$ is the angular frequency of the incoming coherent radiation, $t$ is time, $\hbar$ is the reduced Planck constant, $\mu$ is the transition dipole moment of the QD, $E_0$ is the amplitude of the coherent external illumination $E_f$ (which takes the form $E_f = E_0(e^{- i\omega t}+ e^{+ i\omega t})$, and $c.c.$ denotes the complex conjugate of the respective preceding expression. The absolute permittivity of free space is denoted by $\epsilon_0$ and the permittivities of the QD material and the submerging medium are denoted by $\epsilon_\text{s}$ and $\epsilon_\text{b}$, respectively. The screening of the field incident on/emanated by the exciton due to the QD material is captured using the screening factor $\epsilon_\text{effS} = (2\epsilon_\text{b}+\epsilon_\text{s})/(3\epsilon_\text{b})$. The Rabi frequency of the QD TLA which has been modified due to the plasmonic impact is denoted by $\Omega^r$. The orientation parameter $s_\alpha = -1$ when the QD transition dipole is parallel to the MNP surface (perpendicular to the MNP-QD axis), and $s_\alpha = 2$ when the QD transition dipole is perpendicular to the MNP surface (along the MNP-QD axis). The slowly time-varying amplitude of the off-diagonal QD density matrix element $\rho_\text{eg}$ is denoted by $\tilde{\rho}_\text{eg}$ (where $\rho_\text{eg} = \tilde{\rho}_\text{eg} e^{-i\omega t}$).

Notice that the excitonic Hamiltonian in equation (\ref{Eq:QD_Hamiltonian}), when taken in isolation, describes a closed quantum system where the impact of the submerging environment is not yet taken into consideration. The exciton interacts with its surroundings forming an open quantum system that exhibits irreversible dynamics that can be accounted for using Lindblad terms in the master equation of the QD density matrix $\hat{\rho}$ as follows \cite{hapuarachchi2020influence, hapuarachchi2018exciton, artuso2012optical},
\begin{equation}\label{Eq:Master_Equation}
\dot{\hat{\rho}} = \frac{i}{\hbar}\big[\hat{\rho}, \hat{\mathcal{H}}\big] + \lambda_1\mathcal{L}_{\hat{\sigma}^-}+
\lambda_2\mathcal{L}_{\hat{\sigma}^+}+
\lambda_3\mathcal{L}_{\hat{\sigma}^+\hat{\sigma}^-}.
\end{equation}
The three Lindblad terms $\lambda_1\mathcal{L}_{\hat{\sigma}^-}$, $\lambda_2\mathcal{L}_{\hat{\sigma}^+}$ and $\lambda_3\mathcal{L}_{\hat{\sigma}^+\hat{\sigma}^-}$ represent bath induced decay of the two level atomic system from the excited to ground state, bath induced excitation from ground to excited state, and elastic scattering processes between the bath and the quantum system, respectively. Their expansion takes the form,
\begin{equation}
\mathcal{L}_{\hat{A}} = 2\hat{A}\hat{\rho}\hat{A}^\dagger - \hat{A}^\dagger\hat{A}\hat{\rho} - \hat{\rho}\hat{A}^\dagger\hat{A}.
\end{equation}
For optical frequencies of our interest, even near room temperature, $\lambda_2\approx 0$ \cite{artuso2012optical}. Let us define the population relaxation time $\tau^\measuredangle$ ($\measuredangle=\lbrace\perp, \parallel\rbrace$) of the QD exciton that leads to a mixing between populations or the diagonal density matrix elements $\rho_\text{gg}$ and $\rho_\text{ee}$ \cite{artuso2012optical}, and the corresponding dephasing time $T^\measuredangle$ which causes losses in the off diagonal density matrix elements \cite{sadeghi2010tunable,kosionis2012nonlocal} using the relations $\tau^\measuredangle = 1/(2\lambda_1)$ and $T^\measuredangle = 1\big/(\lambda_1 + \lambda_3)$ \cite{artuso2012optical}. By rearranging these equations and setting $\lambda_3 = 1/t_p$ where $t_p$ is the pure dephasing time, we can obtain \cite{graham2011pure}, 
\begin{equation}\label{Eq:Dephasing_time}
	T^\measuredangle = \frac{2 \tau^\measuredangle t_p}{t_p + 2 \tau^\measuredangle}.
\end{equation}

With the aid of above definitions, the matrix form of the master equation (\ref{Eq:Master_Equation}) in the basis space $\{ |g\rangle , |e\rangle \}$ is obtained as \cite{hapuarachchi2020influence, artuso2012optical},

\begin{equation}\label{Eq:Master_equation_matrix_form}
\dot{\hat{\rho}} = \frac{i}{\hbar}\begin{bmatrix}
-\mu E_\text{qd}(\rho_\text{ge}-\rho_\text{eg}) &  -\mu E_\text{qd}(\rho_\text{gg}-\rho_\text{ee})+\hbar\omega_0\rho_\text{ge} \\
-\mu E_\text{qd}(\rho_\text{ee}-\rho_\text{gg})-\hbar\omega_0\rho_\text{eg} & -\mu E_\text{qd}(\rho_\text{eg}-\rho_\text{ge}) \\
\end{bmatrix} -
\begin{bmatrix}
(\rho_\text{gg}-1)\big/{\tau^\measuredangle} & \rho_\text{ge}\big/T^\measuredangle\\
\rho_\text{eg}\big/T^\measuredangle & {\rho_\text{ee}}\big/{\tau^\measuredangle}
\end{bmatrix}.
\end{equation}

By element-wise comparison of the left and right hand sides of (\ref{Eq:Master_equation_matrix_form}), we can obtain the following Bloch equations for the two-level excitonic system embedded in the QD,
\begin{subequations} \label{Eq:Bloch_equations}
	\begin{align}
	\dot{\rho}_\text{ee} &= -\frac{\rho_\text{ee}}{\tau^\measuredangle} + i\Omega^r\tilde{\rho}_\text{ge} - i\Omega^{r*}\tilde{\rho}_\text{eg} \label{Eq:rhoee_tilde_dot},\\
	\dot{\rho}_\text{gg} &= \frac{\rho_\text{ee}}{\tau^\measuredangle} - i\Omega^r\tilde{\rho}_\text{ge} + i\Omega^{r*}\tilde{\rho}_\text{eg} \label{Eq:rhogg_tilde_dot},\\
	\dot{\tilde{\rho}}_\text{eg} &=-\left[ i(\omega_0 - \omega) + 1\big/T^\measuredangle\right]\tilde{\rho}_\text{eg} + i\Omega^r(\rho_\text{gg} - \rho_\text{ee}), \label{Eq:rhoeg_tilde_dot}
	\end{align}
\end{subequations}
By solving these complex coupled differential equations for the steady-state (by setting $\dot{\rho}_\text{ee} = \dot{\rho}_\text{gg} = \dot{\tilde{\rho}}_\text{eg} = 0$) or temporally, we can analyse emission behaviour of the QD in the presence of the MNP. The same set of equations can be used to analyse the emission behaviour of the isolated QD by setting $z_0\to \infty$ in (\ref{Eq:E_qd_tilde_plus}), which converts $\Omega^r$ to $\Omega_0 = \mu E_0/(\hbar\epsilon_\text{effS})$. This yields the following simple steady-state expression for the excited state population $\left[\rho_\text{ee}\right]_\text{ref}$ of the isolated QD,
\begin{equation}\label{Eq:SS_rho_ee}
	\left[\rho_\text{ee}\right]_\text{ref} = \frac{2 \tau_\text{ref} T_\text{ref} \Omega_0^2}{T_\text{ref}^2(\omega-\omega_0)^2 + 4 \tau_\text{ref} T_\text{ref} \Omega_0^2 + 1},
\end{equation}
where $\tau_\text{ref}$ is the decay time of the isolated QD and $T_\text{ref}$ is the respective dephasing time.

Finally, the following parameters were used with the formalism outlined above to obtain the steady state and temporal emission plots presented in the main text: $\epsilon_\text{b}\approx 1.78$, $z_0\sim \SI{17}{\nano\meter}$, external field amplitude $E_0\approx \SI{1.3e4}{\V\per\meter}$ (switched off at $\SI{100}{\nano\second}$ when obtaining the temporal plots), relative permittivity of the QD material $\epsilon_\text{s}\approx 5.8$ \cite{wakaoka2014confined}, decay rates of the isolated QDs $\tau_\text{ref}^\text{Q550}\approx\SI{9.9}{\nano\second}$ and $\tau_\text{ref}^\text{Q610}\approx\SI{12.9}{\nano\second}$, pure dephasing times $t_p^\text{Q550}\sim \SI{14}{\femto\second}$ and $t_p^\text{Q610}\sim \SI{26}{\femto\second}$ (obtained by fitting the linewidths of the experimental emission spectra for the isolated QDs, allowing for $\sim 10\%$ spectral broadening due to dimensional variations across the ensemble), QD radii $r_\text{Q550}\approx\SI{2.25}{\nano\meter}$ and $r_\text{Q610}\approx\SI{3.25}{\nano\meter}$. QD transition dipole moments are approximated (to first order) as $\mu_\text{Q550}\sim r_\text{Q550} e$ and $\mu_\text{Q610}\sim r_\text{Q610} e$ (where $e$ is the elementary charge). AuNS radius $r_\text{m}\sim \SI{5}{\nano\meter}$, bulk plasma frequency $\hbar\omega_\text{p} = \SI{9.02}{\electronvolt}$, bulk damping rate $\hbar\gamma = \SI{0.071}{\electronvolt}$ Fermi velocity $v_f = \SI{1.39e6}{\meter\per\second}$ and electron diffusion constant for gold $\text{D}\approx\SI{8.62e-4}{\meter\squared\per\second}$ \cite{raza2015nonlocal}. The experimentally measured bulk dielectric data $\epsilon_{\text{expt}}$ for gold were obtained using the tabulations by Johnson and Christy \cite{johnson1972optical}.


\bibliography{HybridAssemblies_Anum}